\documentclass[final,3p,euclid]{elsarticle}

\usepackage{amssymb}
\RequirePackage[l2tabu,orthodox]{nag}
\usepackage{microtype}
\usepackage{amssymb}
\usepackage{graphicx}
\usepackage{subfigure}
\usepackage{amsmath}
\allowdisplaybreaks[4]
\biboptions{sort&compress}
\usepackage{amsfonts}
\usepackage{amsbsy}
\usepackage{bm}
\usepackage{array}
\usepackage{times}
\usepackage{color}
\usepackage{ulem}
\usepackage{lineno}
\usepackage{multirow}
\usepackage{mathrsfs}
\usepackage{booktabs}
\usepackage{hyperref}
\hypersetup{colorlinks,linkcolor=red,anchorcolor=blue,citecolor=green}
\usepackage{cleveref}
\usepackage{booktabs}
\usepackage{siunitx}
\usepackage{fmtcount}
\usepackage[noend]{algpseudocode}
\usepackage{algorithmicx,algorithm}
\usepackage{pifont}
\usepackage{tikz}
\usetikzlibrary{shapes.geometric, arrows}
\usepackage{lineno}

\begin{document}
	\nolinenumbers
	\textcolor{red}{This manuscript has been published in Mechanical Systems and Signal Processing. The DOI of this manuscript is: https://doi.org/10.1016/j.ymssp.2022.109775. Please cite as: C. Ding, C. Dang, M. A. Valdebenito, M. G. Faes, M. Broggi, M. Beer, First-passage probability estimation of high-dimensional nonlinear stochastic dynamic systems by a fractional moments-based mixture distribution approach, Mechanical Systems and Signal Processing 185 (2023) 109775. }
	\begin{frontmatter}
		
		
		
		\title{First-passage probability estimation of high-dimensional nonlinear stochastic dynamic systems by a fractional moments-based mixture distribution approach}
		
		
		\author[label1]{Chen Ding}
		\address[label1]{ Institute for Risk and Reliability, Leibniz University Hannover, Callinstr. 34, Hannover 30167, Germany}
		
		\author[label1]{Chao Dang\corref{cor1}}
		\ead{chao.dang@irz.uni-hannover.de}
		
		\author[label2]{Marcos A. Valdebenito}
        \address[label2]{Faculty of Engineering and Sciences, Universidad Adolfo Ibáñez, Av. Padre Hurtado 750, 2562340 Viña del Mar, Chile}
        
        \author[label3]{Matthias G.R. Faes}
        \address[label3]{Chair for Reliability Engineering, TU Dortmund University,  Leonhard-Euler-Str. 5,  Dortmund 44227, Germany}
        
        \author[label1]{Matteo Broggi}

		\author[label1,label4,label5]{Michael Beer}
		\address[label4]{Institute for Risk and Uncertainty, University of Liverpool, Peach Street, Liverpool L69 7ZF, United Kingdom}
		\address[label5]{International Joint Research Center for Resilient Infrastructure \& International Joint Research Center for Engineering Reliability and Stochastic Mechanics, Tongji University, Shanghai 200092, PR China}
		
		
		\cortext[cor1]{Corresponding author}
		
		\begin{abstract}
First-passage probability estimation of high-dimensional nonlinear stochastic dynamic systems is a significant task to be solved in many science and engineering fields, but remains still an open challenge.
The present paper develops a novel approach, termed ‘fractional moments-based mixture distribution’, to address such challenge. 
This approach is implemented by capturing the extreme value distribution (EVD) of the system response with the concepts of fractional moment and mixture distribution. 
In our context, the fractional moment itself is by definition a high-dimensional integral with a complicated integrand. 
To efficiently compute the fractional moments, a parallel adaptive sampling scheme that allows for sample size extension is developed using the refined Latinized stratified sampling (RLSS).
In this manner, both variance reduction and parallel computing are possible for evaluating the fractional moments.
From the knowledge of low-order fractional moments, the EVD of interest is then expected to be reconstructed.
Based on introducing an extended inverse Gaussian distribution and a log extended skew-normal distribution, one flexible mixture distribution model is proposed, where its fractional moments are derived in analytic form.
By fitting a set of fractional moments, the EVD can be recovered via the proposed mixture model.
Accordingly, the first-passage probabilities under different thresholds can be obtained from the recovered EVD straightforwardly. 
The performance of the proposed method is verified by three examples consisting of two test examples and one engineering problem. 

\end{abstract}


\end{frontmatter}

\nolinenumbers

\section{Introduction}\label{}
Stochastic dynamic systems which involve the randomness in internal system properties and/or external dynamic loads are \textcolor{black}{widespread} in various science and engineering fields, such as meteorology, quantum optics, circuit theory and structural engineering \cite{Fokker-Planck1989}.
To assess the effects of input randomness on the system performance, dynamic reliability analysis has drawn increasing attention. 
Generally, dynamic reliability analysis for stochastic dynamic systems can be classified as the first-passage probability evaluation and the fatigue failure probability estimation \cite{lyu2021first}. 
In the literature, the first-passage probability evaluation has been extensively studied over the past several decades.
However, finding efficient and accurate solutions to the first-passage problem still remains challenging.
The reason is twofold:
\textcolor{black}{(1)} the high-dimensional \textcolor{black}{input} randomness and \textcolor{black}{strongly nonlinear behavior} of stochastic dynamic systems may be encountered simultaneously; 
\textcolor{black}{(2)} the first-passage probabilities of such systems under certain thresholds may be relatively small.

The existing approaches for first-passage probability estimation can be broadly divided into four kinds: the out-crossing rate approaches, the diffusion process approaches, the stochastic simulation approaches and the extreme value distribution (EVD) estimation approaches. 
For the out-crossing rate approaches, the first-passage probability is evaluated \textcolor{black}{considering} the time of out-crossing within a time duration on the basis of Rice's formula \cite{rice1945mathematical, naess2008monte, he2009approximation, lutes2012amplitude}.
Such approaches are based on the Poisson assumption that level-crossing events are mutually independent and each happens at most once, or the Markovian assumption that the next crossing event only relates to the present event \cite{li2009stochastic}. 
Although these solutions can be accurate in some special cases, they may be not applicable for general cases. 
Besides, it is hard to derive the joint probability density function (PDF) and its derivatives of the system response of interest when complicated nonlinear stochastic dynamic systems are encountered.
The diffusion process approaches evaluate the first-passage probability by solving a partial differential equation, such as the Kolmogorov backward equation \cite{vanvinckenroye2019reliability} or the Fokker Planck equation \cite{kovaleva2009exact}. 
Solutions to such equations could be derived via the path integration method \cite{iourtchenko2008reliability, kougioumtzoglou2013response, bucher2016first}, stochastic average technique \cite{spanos2014galerkin, dos2019hilbert}, ensemble-evolving-based generalized density evolution equation \cite{lyu2021first,li2012advances}, etc. 
Nevertheless, this kind of approach is mostly applicable for nonlinear stochastic dynamic systems enforced by white noise. 
For the stochastic simulation approach, the extensively used Monte Carlo simulation (MCS) \cite{shinozuka1972monte} is able to address problems regardless of their dimensions and nonlinearities. 
However, MCS is inefficient and even infeasible to assess a small probability for an expensive-to-evaluate model since a considerably large number of simulations are required. 
Although some variants of MCS have been developed, such as important sampling \cite{au2003important, wang2016cross,kanjilal2021cross,zhao2022efficient} and subset simulation \cite{au2001estimation,ching2005hybrid,au2007application}, they still suffer their respective limitations concerning efficiency, accuracy and applicability, etc.

\textcolor{black}{Recently, the EVD estimation approaches have attracted lots of attention.
This is because} 
once the EVD of system response of interest is obtained, the first-passage probability can be straightforwardly and conveniently evaluated \cite{chen2007extreme}. 
Nevertheless, the analytical solution to the EVD is difficult and even impossible to be obtained for a general nonlinear stochastic dynamic system.
Therefore, various approximation methods have been developed to estimate the EVD, which can be roughly classified as probability conservation-based methods and moments-based methods.
According to the principle of probability conservation, the probability density evolution method (PDEM) \cite{li2009stochastic,chen2007extreme} and direct probability integral method (DPIM) \cite{chen2021unified} are derived, which can be used for the purpose of EVD estimation. 
However, since such methods are typically dependent on the partition of random variable space, their application for high-dimensional problems may be challenging. 
Moment-based methods, on the other hand, estimate the first-passage probability by fitting an appropriate parametric distribution model to the EVD, and the free parameters of the distribution model are obtained from the estimated moments of the EVD. 
The integer moments-based methods can be adopted to recover the EVD \cite{he2016estimate,ZHAO2022108967}, where high-order integer moments, i.e., skewness and kurtosis, need to be considered. 
Yet it is difficult to evaluate such high-order integer moments using a small sample size, due to their large variability \cite{li2019improved}. 
To alleviate such difficulty, a series of methods based on non-integer moments, such as fractional moments and linear moments, have been developed. 
The fractional moments-based maximum entropy methods \cite{xu2016efficient,xu2018two,xu2019novel,chen2020efficient} can estimate the first-passage probabilities of nonlinear stochastic dynamic systems from low to high dimensions. 
\textcolor{black}{However, it is difficult to solve the non-convex optimization problem that is typically encountered, and the obtained results can be easily trapped into local optimum.}
Besides, due to the polynomials involved in the maximum entropy density, the recovered EVD can have unexpected oscillating distribution tail, which then leads to \textcolor{black}{an inaccurate evaluation} of the first-passage probability. 
Two mixture parametric distribution methods in conjunction with fractional moments \cite{dang2020mixture} or moment-generating function \cite{dang2021approach} are developed.
These methods enable to evaluate first-passage probabilities of high-dimensional and strongly nonlinear stochastic dynamic systems from a small number of simulations.
Furthermore, a fractional moments-based shifted generalized lognormal distribution method \cite{chen2021seismic} is utilized to assess seismic reliability of a practical bridge subjected to spatial variate ground motions.  
Besides, the linear moments-based polynomial normal transformation distribution method \cite{zhang2021dynamic} 
is developed to analyze high-dimensional dynamic systems with deterministic structural parameters subjected to stochastic excitations. 

Overall, the fractional moments-based methods offer the possibility to deal with both high-dimensional and strongly nonlinear stochastic dynamic systems from a reduced number of simulations, even with small first-passage probabilities.
In view of this, the present paper mainly focuses on such methods. 
Despite those attractive features, the fractional moments-based methods still have two main problems to be solved.
On one hand, the sample size for evaluating fractional moments is usually empirically fixed.
This is primarily because the sampling-based schemes adopted by the existing methods do not allow for the sample size extension. 
However, the optimal sample size should be problem-dependent. 
With a predetermined sample size, the adopted sampling methods may encounter over-sampling or under-sampling, leading to a waste of over-all computational efforts or unsatisfactory accuracy of estimated fractional moments.
On the other hand, the success of fractional moments-based methods for first-passage probability evaluation also depends on the selection of an appropriate distribution model.
Although the existing distribution models are capable of representing EVDs for some problems, their flexibility and applicability are limited.
Hence, for a wide range of problems, they \textcolor{black}{may} still lack the ability to accurately recover the EVDs over the entire distribution domain, especially for the tails. 

In this paper, we propose a fractional moments-based mixture distribution approach to estimate the first-passage probabilities of high-dimensional and strongly nonlinear stochastic dynamic systems. 
It is worth mentioning that the randomness from both internal system properties and external excitations is taken into account. 
\textcolor{black}{The main contributions of this study are summarized as follows. }
First, a parallel adaptive sampling scheme is proposed for estimating the fractional moments, as opposed to the traditional fixed sample size scheme. 
Such a new scheme enables to extend the sample size sequentially, i.e., one at a time or several at a time. 
The optimal sample size for fractional moment estimation is determined by introducing a convergence criterion. 
In fact, a sequential sampling method with the ability to effectively reduce variance in high-dimensional problems, named Refined Latinized stratified sampling (RLSS) \cite{shields2016refined}, is suitable for achieving our purposes and is employed within the proposed scheme. 
Second, one novel and versatile mixture distribution model is proposed to reconstruct the EVD with the knowledge of its estimated fractional moments.
This model is based on the extension of the conventional inverse Gaussian distribution and the log transformation of the extended skew-normal distribution. 
The analytical expression of the fractional moments for such mixture distribution is derived, and a fractional moments-based parameter estimation technique is developed. 

The remainder of this paper is organized as follows. Section \ref{section:section2} outlines the first-passage probability estimation of a stochastic dynamic system from the perspective of EVD. In section \ref{section:section3}, the proposed fractional moments-based mixture distribution approach is described in detail, including a parallel adaptive scheme for fractional moments evaluation and a flexible mixture distribution model for EVD reconstruction.
Three examples are given in section \ref{section:section4} to demonstrate the performance of the proposed method. The paper is closed with some concluding remarks in section \ref{section:section5}.

\section{First-passage probability estimation of stochastic dynamic systems} \label{section:section2}

\subsection{Stochastic dynamic systems}
Consider a stochastic dynamic system that is governed by the following state-space equation:
\begin{equation}\label{eq:state_equation}
	\mathbf{\dot{Y}}\left( t \right) =\mathbf{Q}\left( \mathbf{Y}\left( t \right) ,\mathbf{U},t \right), 
\end{equation}
with an initial condition
\begin{equation}
\mathbf{{Y}}\left( 0 \right) = \textcolor{black}{\boldsymbol{y}_0},
\end{equation}
where $\mathbf{{Y}} = \left( Y_1, Y_2, ... , Y_{n_d} \right)$ is a $n_d$-dimensional state vector; $\mathbf{Q} = \left( Q_1, Q_2, ... , Q_{n_d} \right)$ is a dynamics operator vector; $\mathbf{U} = \left( U_1, U_2, ... , U_{n_s} \right)$ is a $n_s$-dimensional random parameter vector with a known joint probability density function (PDF) $p_{\mathbf{U}} \left( \mathbf{u} \right)$; $\mathbf{u} = \left( u_1, u_2, ... , u_{n_s} \right)$ denotes a realization of $\mathbf{U}$; $t$ denotes the time. 
Note that Eq. (\ref{eq:state_equation}) can be strongly nonlinear, which may be caused by material, geometrical, \textcolor{black}{or contact nonlinearities} inherent in the stochastic dynamic system.
In addition, hundreds or thousands of random variables can be included in the vector ${\mathbf{U}}$ due to the randomness from system properties and external excitations.

For a well-posed stochastic dynamic system, the solution to Eq. (\ref{eq:state_equation}) is unique and depends on the vector $\mathbf{U}$, which can be assumed to be: 
\begin{equation}
 \left[ \mathbf{Y}\left( t \right), \mathbf{\dot{Y}}\left( t \right) \right] = \left[ \mathbf{H}_{\mathbf{Y}}\left( \mathbf{U},t \right) , \frac{\partial \mathbf{H}_{\mathbf{Y}}\left( \mathbf{U},t \right)}{\partial t}  \right],
\end{equation}  
where $\mathbf{H_{Y}}$ and $\frac{\partial \mathbf{H}_{\mathbf{Y}}}{\partial t}$ are the deterministic operators. 

If we consider the system responses of interest for reliability analysis, say $\boldsymbol{\mathcal{W}}\left( t \right)= \left\{\mathcal{W}_1\left( t \right), \mathcal{W}_2\left( t \right), ..., \mathcal{W}_{n_d}\left( t \right) \right\}$, they can be evaluated from their relations to the state vectors:
\begin{equation}\label{eq:state_system_response}
	{\boldsymbol{\mathcal{W}}\left( t \right)}=\boldsymbol{\varPsi} \left[ \mathbf{Y}\left( t \right) , \mathbf{\dot{Y}}\left( t \right) \right] = \mathscr{H}{\left( \mathbf{U},t \right)},
\end{equation}  
where $\boldsymbol{\varPsi}$ is the transfer operator; and $\mathscr{H}$ denotes the mapping relation from $\mathbf{U}$ and $t$ to $\boldsymbol{\mathcal{W}}\left( t \right)$. 
Accordingly, the $q$-th component of \textcolor{black}{$\boldsymbol{\mathcal{W}}\left( t \right)$} is denoted by ${{\mathcal{W}_q}\left( t \right)} = \mathscr{H}_q{\left( \mathbf{U},t \right)}, q=1,...,n_d$. 
For notational simplicity, the subscript $q$ is omitted hereafter, and only a component $\mathcal{W}\left( t \right)$ is considered in the following.

\subsection{First-passage probability estimation by EVD}
For a stochastic dynamic system, the first-passage probability is the probability that the system response of interest exceeds a certain safe domain for the first time within a given time range. Accordingly, assuming $T$ is the time duration, we have 
\begin{equation}
P_f=\mathrm{Pr}\left\{ \mathcal{W} \left( t \right) \notin \varOmega _{\mathrm{safe}},\exists t\in \left[ 0,\left. T \right] \right. \right\}, 
\end{equation}
where ${P_f}$ is first-passage probability; $\mathrm{Pr}$ is probability operator; $\varOmega _{\mathrm{safe}}$ denotes the safe domain. 
\textcolor{black}{According to different application backgrounds, the boundary of $\varOmega _{\mathrm{safe}}$ can be different, such as one boundary, double boundary, and circle boundary \cite{li2009stochastic}.}
In the case of symmetric double boundary problem, the first-passage probability can be further written as: 
\begin{equation}\label{eq:First-passage_probability}
	P_f=\mathrm{Pr}\left\{ \left| \mathcal{W}\left( t \right) \right|>b_{\lim}, \exists t\in \left[ 0,\left. T \right] \right. \right\}, 
\end{equation}
where $b_{\lim}$ is the given threshold \textcolor{black}{that limits the symmetric bounds of $\varOmega _{\mathrm{safe}}$, and $\left|\cdot\right|$ is the absolute value operator}. 
In the present study, the first-passage probability defined by Eq. (\ref{eq:First-passage_probability}) is of concern.

\textcolor{black}{Note that if the system response in the time period $\left[ 0,\left. T \right] \right.$ remains below the boundary of $\varOmega _{\mathrm{safe}}$, the first-passage probability will be equal to zero. 
From this perspective, once the extreme value of system response exceeds the boundary, the system fails. 
Accordingly, Eq. (\ref{eq:First-passage_probability}) can be rewritten as 
\begin{equation}\label{eq:MEVD1}
P_f=\mathrm{Pr}\left\{ \max \left\{ \left| \mathcal{W} \left( t \right) \right| \right\} >b_{\lim},\forall t\in \left[ 0,\left. T \right] \right. \right\} 
\\
=\mathrm{Pr}\left\{ \mathcal{Z} >b_{\lim} \right\},
\end{equation}
where $\mathcal{Z}=\mathop {\mathrm{max}} \limits_{t\in \left[ 0,\left. T \right] \right.}\left\{ \left| \mathcal{W}\left( t \right) \right| \right\}$. 
}
Note that $\mathcal{Z}$ is always positive, and depends on the random parameter vector $\mathbf{U}$.
If we denote the functional relationship between $\mathcal{Z}$ and $\mathbf{U}$ as ${G}$, then \textcolor{black}{we have $\mathcal{Z}= G \left( {\mathbf{U}} \right)$ and $P_f=\mathrm{Pr}\left\{ \mathcal{Z}= G \left( {\mathbf{U}} \right) > b_{\lim} \right\}$.}

According to classical probability theory, once the probability distribution of $\mathcal{Z}$, which is also referred to as \textcolor{black}{extreme value distribution (EVD)}, is obtained, Eq. (\ref{eq:MEVD1}) can be straightforwardly calculated from the EVD.
Let ${f_{\mathcal{Z}}\left( z \right)}$ and $F_{\mathcal{Z}}\left( z \right)$ be the PDF and cumulative distribution function (CDF) of $\mathcal{Z}$. 
Then the first-passage probability reads
\begin{equation}\label{eq:MEVD2}
	P_f=\int_{b_{\lim}}^{+\infty}{f_{\mathcal{Z}}\left( z \right)}\mathrm{d}z=1-F_{\mathcal{Z}}\left( {b_{\lim}} \right). 
\end{equation}

It should be pointed out that the first-passage probability is easy to be obtained from Eq. (\ref{eq:MEVD2}) once the PDF or CDF of $\mathcal{Z}$ is known. 
However, how to estimate the EVD \textcolor{black}{of $\mathcal{Z}$} is quite challenging. 
This is because deriving an analytical expression for the EVD is intractable even for some simple stochastic \textcolor{black}{responses}, not to mention the stochastic \textcolor{black}{responses} of high-dimensional and strong-nonlinear stochastic dynamic systems. 
Therefore, to tackle such challenge, an EVD estimation method is proposed in the following section. 

\noindent \textcolor{black}{\textbf{Remark 1.} 
For system failure probability evaluation, the above-mentioned EVD estimation method can also be applied by using the theory of equivalent extreme-value events \cite{li2007equivalent}. Briefly speaking, the system failure can be regarded as a compound event of
multiple random events, where a single random event can be described by an inequality associated with a single response and its threshold. 
Based on the inequality relationship between the involved random events, the compound event can be equated to an equivalent extreme-value event whose threshold can be obtained by a linear combination of the thresholds of the involved random events. 
In this manner, the system failure probability can be assessed by the Eq. (\ref{eq:MEVD2}), where $\mathcal{Z}$ is the equivalent extreme-value event.
The interested readers can refer to Ref. \cite{li2007equivalent} for more details.}

\section{A fractional moments-based mixture distribution approach}\label{section:section3}
In this section, we propose a novel fractional moments-based mixture distribution approach to approximate the EVD in an efficient and accurate way. 
The proposed method consists of two main parts. 
First, a parallel adaptive scheme is proposed for fractional moments estimation, which allows sequential sample size extension until a prescribed convergence criterion is satisfied. 
Second, from the knowledge of estimated fractional moments, an eight-parameter mixture distribution model with increased flexibility is developed to capture the main body and distribution tail of the EVD. 

\subsection{Characterizing EVD by fractional moments}
The analytical expression of EVD can not be directly obtained for a general high-dimensional and nonlinear stochastic dynamic system, as discussed earlier. 
To this end, we have to resort to some indirect methods that can approximate the EVD from a limited number of sample data. The fractional moment, as a generalization of the traditional integer moment, has received a growing interest to characterize a positive random variable in many fields.  
More recently, it has also been introduced to the area of EVD characterization \cite{xu2019novel, chen2020efficient, dang2020mixture, chen2021seismic}. 

\subsubsection{Concept \textcolor{black}{and properties} of fractional moments}\label{section:concept_FM}
The $r$-th fractional moment of the positive random variable $\mathcal{Z}$ is defined as \cite{dang2020mixture}
\begin{equation}\label{eq:frac_mom_def}
M_{\mathcal{Z}}^{r}=E\left[ \mathcal{Z} ^r \right] =\int_0^{+\infty}{z^rf_{\mathcal{Z}}\left( z \right) \mathrm{d}z},
\end{equation}
where $r$ can be any real number and $E\left[\cdot\right]$ denotes the expectation operator. 
Note that when $r$ takes an integer value, Eq. (\ref{eq:frac_mom_def}) yields the $r$-th integer moment of $\mathcal{Z}$. 
Therefore, for any positive random variable, the integer moment of the variable is a special case of its fractional moment. 

If one expands $\mathcal{Z}^r$ around its mean value $\mu_{\mathcal{Z}}=M_{\mathcal{Z}}^{1}$ using the Taylor series expansion\textcolor{black}{, we have}
\begin{equation}\label{eq:frac_mom1}
	\mathcal{Z}^r=\sum_{k=0}^{\infty}{\binom{r}{k} \mu_{\mathcal{Z}}^{r-k}\left( z-\mu_{\mathcal{Z}} \right) ^k},
\end{equation}
where the fractional binomial coefficient $\binom{r}{k}$ can be computed as $\binom{r}{k} = \frac{r\left( r-1 \right) \cdots \left( r-k+1 \right)}{k\left( k-1 \right) \cdots 1}$, and $k$ can be any non-negative integer. 
\textcolor{black}{Taking} the expectation of both sides of Eq. (\ref{eq:frac_mom1}) \textcolor{black}{yields}:
\begin{equation}\label{eq:frac_mom2}
	E\left[ \mathcal{Z}^r \right] =\sum_{k=0}^\infty {\binom{r}{k} \mu _{\mathcal{Z}}^{r-k}E\left[ \left( z-\mu _{\mathcal{Z}} \right) ^k \right]}.
\end{equation}
It can be seen that the \textcolor{black}{right-hand} side of Eq. (\ref{eq:frac_mom2}) \textcolor{black}{contains} an infinite number of integer moments, i.e., $E\left[ \left( z-\mu _{\mathcal{Z}} \right) ^k \right]$, and the \textcolor{black}{left-hand} side of Eq. (\ref{eq:frac_mom2}) is exactly the $r$-th fractional moment. 
Hence, Eq. (\ref{eq:frac_mom2}) implies that a single $r$-order fractional moment can embody statistical information of numerous integer moments. 
\textcolor{black}{Further, as observed from Eq. (\ref{eq:frac_mom2}), when $r$ is fixed, the value of coefficient $\binom{r}{k} \mu _{\mathcal{Z}}^{r-k}$ decreases as $k$ increases; when $k$ is fixed, $\binom{r}{k} \mu _{\mathcal{Z}}^{r-k}$ increases as $r$ increases.
This indicates that the higher the fractional order, the greater the contribution of higher-order integer moments.
Since higher-order integer moments can provide more information about the shape of EVD, higher-order fractional moments reflect more statistical features of EVD than lower-order fractional moments.
}
In addition, it should be mentioned that higher-order fractional moments have higher variability and are more difficult to obtain than lower-order fractional moments \cite{dang2020mixture,li2019improved}. 
\textcolor{black}{Note that one is able to generate any number of fractional moments given the range of fractional orders.
However, one can only generate a fixed number of integer moments if the maximum integer order is given. 
As a compromise, a set of fractional moments up to second order, as adopted in Ref. \cite{dang2020mixture}, is used in this work.}

\subsubsection{Parallel adaptive estimation of fractional moments}\label{RLSS}
According to the principle of probability conservation, Eq. (\ref{eq:frac_mom_def}) can be rewritten in the random variable space of $\mathbf{U}$:
\begin{equation}\label{eq:frac_mom_def2}
M_{\mathcal{Z}}^{r}=\int_{\Omega _{\mathbf{U}}}{G^r\left( \mathbf{u} \right) p_{\mathbf{U}}\left( \mathbf{u} \right) \mathrm{d}\mathbf{u}},
\end{equation}
where $\Omega _{\mathbf{U}}$ denotes the random variable space of $\mathbf{U}$.  
For a general stochastic dynamic system, a considerably large number of random variables are collected in $\mathbf{U}$, and strong nonlinearity exists in $G\left( \mathbf{U} \right)$. 
In addition, the expression of $G\left( \mathbf{U} \right)$ cannot be explicitly given.
Hence, a high-dimensional integral with a complex and implicit integrand is involved in Eq. (\ref{eq:frac_mom_def2}), which is impossible to solve analytically. 

Alternatively, we can resort to the sampling methods to approximate the high-dimensional integral involved in Eq. (\ref{eq:frac_mom_def2}). 
In the literature, various variance reduction sampling methods with fixed sample sizes are employed to facilitate the estimation of fractional moments.
Under this setting, $M_{\mathcal{Z}}^{r}$ can be approximated as:
\begin{equation}\label{eq:raw_moments}
\hat{M}_{\mathcal{Z}}^{r}=\sum_{k=1}^N{\varpi _k\cdot G^r\left( \mathbf{u}_k \right)},
\end{equation}
where ${N}$ denotes the sample size; ${\varpi_k}$ represents the $k$-th sample weight, $k=1,...,N$; ${\mathbf{u}}_{k}$ is the $k$-th sample of random variables $\mathbf{U}$.
Note that most variance reduction sampling methods do not allow sample size extension, and thus require ${N}$ to be specified in advance from experience. 
However, for estimating fractional moments, an ``optimal sample size'' is desired, which is problem-dependent, and cannot be known in advance for a specified first-passage problem.
The optimal sample size enables the estimation to \textcolor{black}{strike a} balance between accuracy and computational efficiency.
\textcolor{black}{However,} with a predefined sample size, the fractional moment estimation may lose such balance, and may be trapped into over-sampling or under-sampling situations. 
Specifically, if an overly conservative sample size is pre-specified, i.e., too many samples are taken, oversampling occurs and \textcolor{black}{leads to unnecessary computational waste.} 
On the other hand, if the predefined sample size \textcolor{black}{is too small}, under-sampling takes place, resulting in inaccurate evaluation of the fractional moments.

\begin{figure}[!htb]
    \centering
    \includegraphics[width=0.8\linewidth]{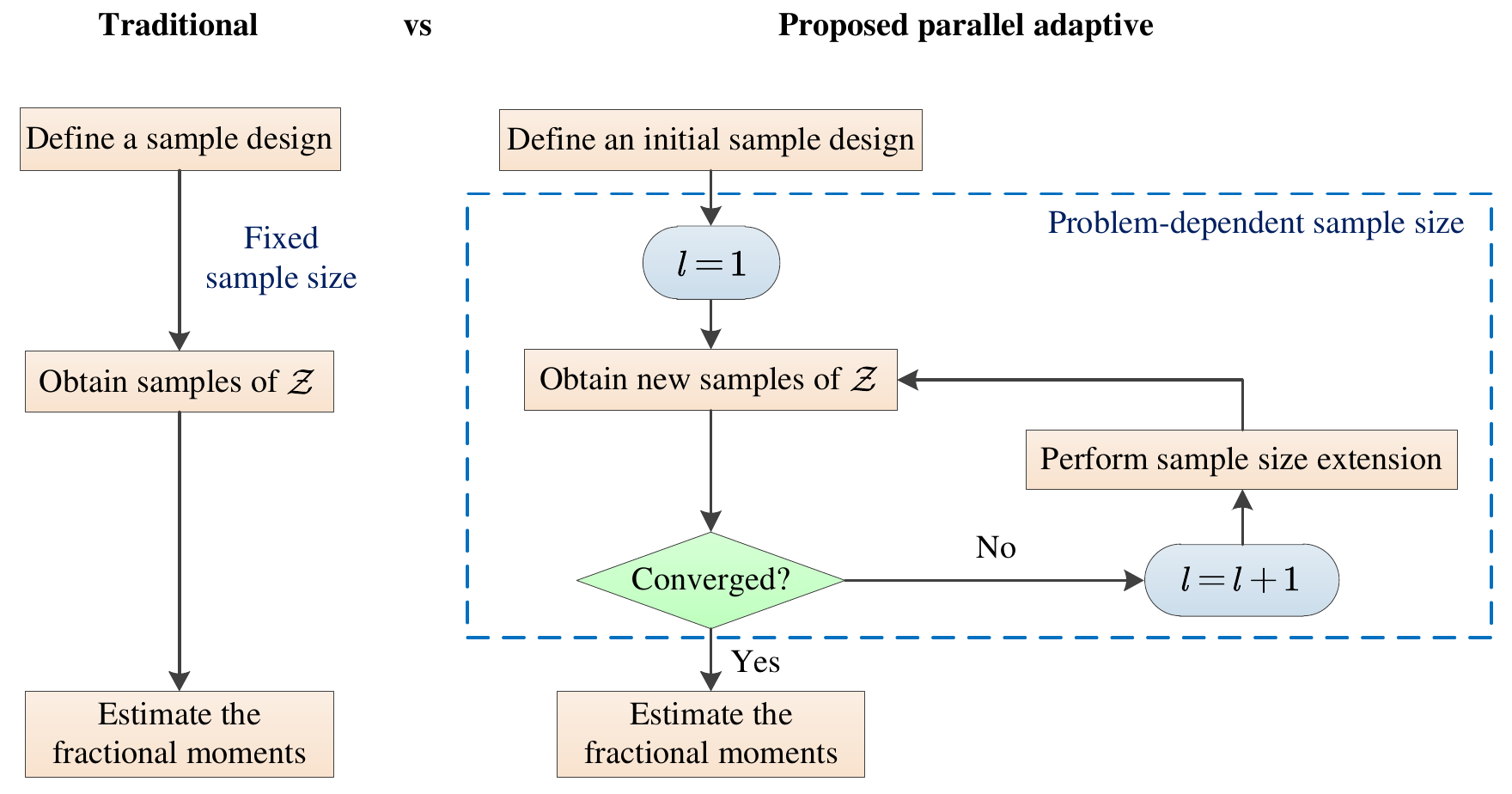}
    \caption{Comparison of traditional sampling scheme and proposed parallel adaptive scheme}
    \label{fig:generalflowchart}
\end{figure}

To tackle with such dilemma, an adaptive sampling scheme should be developed for estimating fractional moments. 
One feasible strategy is to generate samples one at a time or several at a time, and enrich the sample size progressively until a specified convergence criterion is satisfied.
In this manner, sample size extension is allowed, and the sample size can be obtained adapted to different problems, which enables the estimated fractional moments to achieve \textcolor{black}{both the} desired accuracy and computational efficiency.
In addition, parallel computing technique can be equipped to further accelerate the computational speed of such process.
As such, we shall name this sampling scheme as parallel adaptive sampling scheme.
To illustrate the advantages of proposed scheme, Fig. \ref{fig:generalflowchart} shows the comparison between traditional sampling scheme and proposed parallel adaptive sampling scheme.
In this figure, $l$ denotes the $l$-th time of sample size extension, and $l \in {\mathbb{Z}^{+}}$. 
As seen, by the proposed sampling scheme, the sample size for a given first-passage problem can be determined in an adaptive way, where fractional moments can be approximated with a desired accuracy.
In addition, it is quite time-saving to evaluate additional samples of $\mathcal{Z}$ only when it is required. 
In the process of estimating the additional samples of $\mathcal{Z}$, the analysis time can be further decreased by adopting parallel computing technique. 

By employing the proposed parallel adaptive sampling scheme, $\hat{M}_{\mathcal{Z}}^{r}$ after the $l$-th sample size extension can be computed as follows:
\begin{equation}\label{eq:raw_moments_parallel_adaptive}
\hat{M}_{\mathcal{Z}}^{r}=\sum_{k=1}^{\left( l-1 \right) \hbar}{\varpi ^{\left( k \right)}\cdot G^r\left( \mathbf{u}^{\left( k \right)} \right)}+\sum_{k=\left( l-1 \right) \hbar +1}^{l\hbar}{\varpi ^{\left( k \right)}\cdot G^r\left( \mathbf{u}^{\left( k \right)} \right)},
 \end{equation}
where the number of samples added in each time of sample size extension is denoted as $\hbar$ and $\hbar \in {\mathbb{Z}^{+}}$; the current sample size is $l \hbar$; the weight is reallocated in the $l$-th sample size extension and satisfies $\sum_{k=1}^{l \hbar}{\varpi ^{\left( k \right)}}=1$; $\left\{ \mathbf{u}^{\left( \left( l-1 \right) \hbar +1 \right)},...,\mathbf{u}^{\left( l\hbar \right)} \right\} $ are the newly added samples in the $l$-th sample size extension, while $\left\{ \mathbf{u}^{\left( 1 \right)},...,\mathbf{u}^{\left( \left( l-1 \right) \hbar \right)} \right\} $ are samples generated in the previous $\left(l-1\right)$ sample size extensions. 
Note that when $l=1$, initial samples of $\mathcal{Z}$, i.e., $\left\{ G\left( \mathbf{u}^{\left( k \right)} \right) \right\} _{k=1}^{ \hbar}$ are evaluated.
Since $\left\{ G\left( \mathbf{u}^{\left( k \right)} \right) \right\} _{k=1}^{\left( l-1 \right) \hbar}$ have been already obtained in the previous $\left(l-1\right)$ sample size extensions, one only needs to evaluate $\left\{ G\left( \mathbf{u}^{\left( k \right)} \right) \right\} _{k=\left( l-1 \right) \hbar +1}^{l\hbar}$ in the $l$-th sample size extension.

In order to achieve the proposed parallel adaptive sampling scheme, the key is to employ a sampling strategy that allows sequential sample size extension. 
Simple random sampling method, i.e., Monte Carlo simulation (MCS), can naturally meet such aim. 
To obtain a better precision of fractional moments with fewer computational efforts, one can apply a variance reduction sampling method to the proposed sampling scheme. 
In addition, sampling methods that are applicable to high-dimensional problems are also desired. 
In fact, one recently developed sequential stratified sampling technique, termed refined Latinized stratified sampling (RLSS) \cite{shields2016refined}, is suitable for our purposes. 
On one hand, RLSS is advantageous as it owns the ability to achieve effective variance reduction in terms of both main/additive effects and variable interaction that appear in $G\left(\mathbf{U}\right)$. 
On the other hand, RLSS is applicable to problems involving low- and high-dimensional input random variables. 
By using the RLSS technique, we can evaluate $\hat{M}_{\mathcal{Z}}^{r}$ according to Eq. (\ref{eq:raw_moments_parallel_adaptive}). 
Since the samples of RLSS are generated in the $\left[ 0,1 \right]^{n_s}$ hyper-rectangular space, we need to transform the RLSS sample points to the original distribution domain of random variables $\mathbf{U}$.
Denote $\boldsymbol{\hat{\varphi} }^{\left(k\right)}$ and $\varpi^{\left(k\right)}$ to be the $k$-th sample point and corresponding weight obtained by RLSS and $\Gamma$ to be the transformation operator, $\hat{M}_{\mathcal{Z}}^{r}$ by RLSS at the $l$-th sample size extension can be evaluated as:
\begin{equation}\label{eq:raw_moments_RLSS}
\hat{M}_{\mathcal{Z}}^{r}=\sum_{k=1}^{\left( l-1 \right) \hbar}{\varpi ^{\left( k \right)}\cdot G^r\left( \Gamma \left( \boldsymbol{\hat{\varphi}}^{\left( k \right)} \right) \right)}+\sum_{k=\left( l-1 \right) \hbar +1}^{l\hbar}{\varpi ^{\left( k \right)}\cdot G^r\left( \Gamma \left( \boldsymbol{\hat{\varphi}}^{\left( k \right)} \right) \right)}.
\end{equation}

A brief illustration of the RLSS technique is discussed in the following. For more details, the interested readership can refer to  \ref{section:appendix} or Ref. \cite{shields2016refined}.

\begin{figure}[!htb]
	\centering
	\subfigure[]{
		\begin{minipage}{7.0cm}
			\centering
			\includegraphics[scale=0.85]{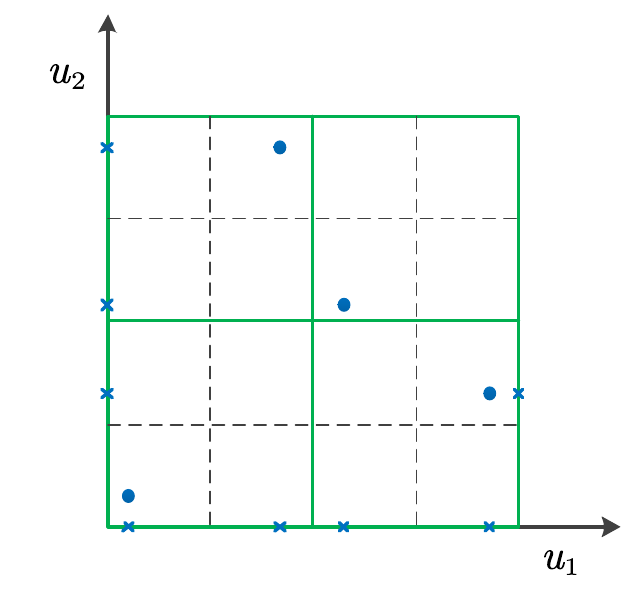}
			\label{fig:illustration of RLSS (a)}
		\end{minipage}
	}%
	\subfigure[]{
		\begin{minipage}{7.0cm}
			\centering
			\includegraphics[scale=0.85]{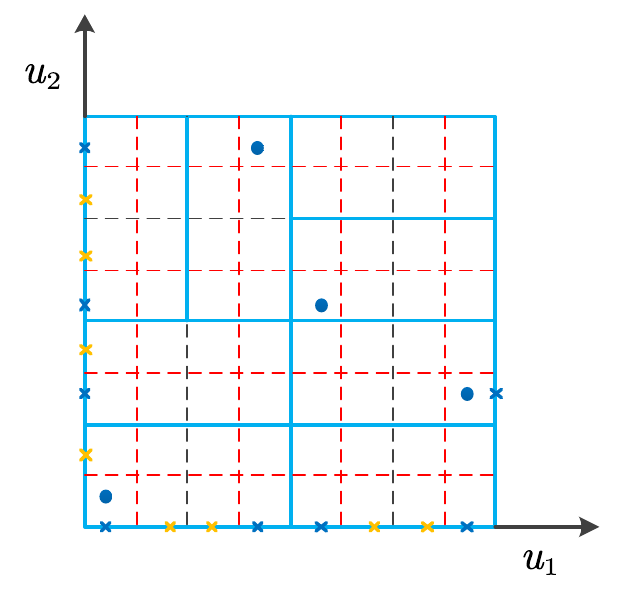}
			\label{fig:illustration of RLSS (b)}
		\end{minipage}
	}%
	
	\subfigure[]{
		\begin{minipage}{7.0cm}
			\centering
			\includegraphics[scale=0.85]{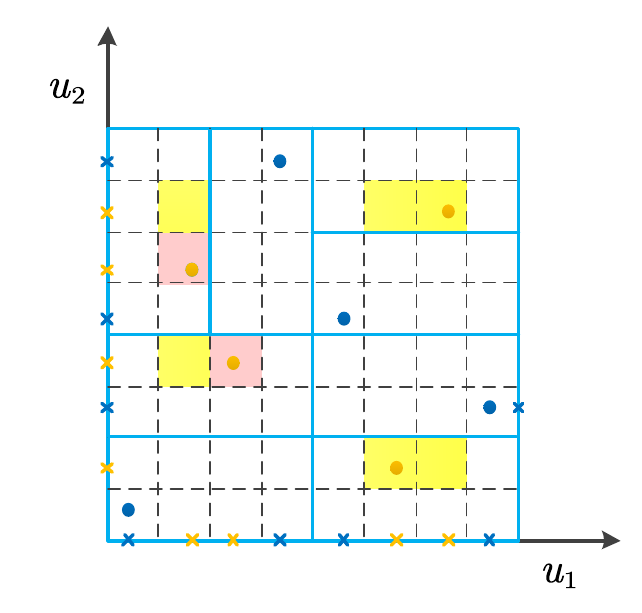}
			\label{fig:illustration of RLSS (c)}
		\end{minipage}
	}%
	\subfigure[]{
		\begin{minipage}{7.0cm}
			\centering
			\includegraphics[scale=0.85]{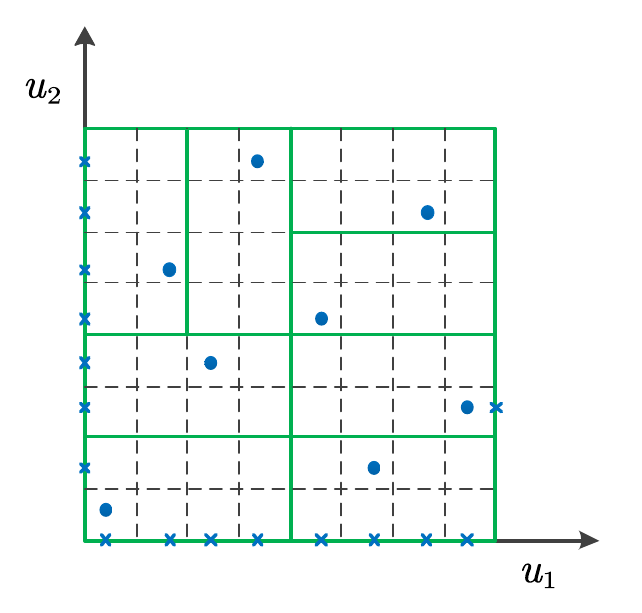}
			\label{fig:illustration of RLSS (d)}
		\end{minipage}
	}%
	\caption{Schematic description of  \textcolor{black}{the} RLSS technique for \textcolor{black}{generating} 8 samples in two dimensions}
	\label{fig:illustration of RLSS}
\end{figure}

The first step of RLSS is generating $\mathcal{N} \ge 1$ samples that follow a so-called Latinized stratifed sampling (LSS) scheme \cite{shields2016generalization}, which implies that these samples fulfill both the properties of Latin hypercube sampling (LHS) and stratified sampling (SS). 
An schematic diagram of a LSS design is shown in Fig. \ref{fig:illustration of RLSS (a)}, considering $\mathcal{N}=4$ and $n_s=2$. 
In this figure, the strata associated with LHS are shown with dashed black line, the strata associated with SS are marked with solid green line, the samples per each dimension of analysis are marked with blue cross marks and the actual samples are marked with blue dots.
It is readily observed that the strata associated with SS possess the same area, 
and boundaries of the strata associated with LHS match those associated with SS, which are the key properties of LSS. 

The second step of RLSS consists of applying a Hierarchical Latin hypercube sampling (HLHS) design \cite{shields2016refined} over the existing LHS design. 
This implies applying a refinement of each LHS strata by subdividing it $\delta$ times, which is illustrated schematically in Fig. \ref{fig:illustration of RLSS (b)}, where $\delta=1$. 
The new strata associated with LHS are shown with red dashed line and the new candidate samples per each dimension on those strata are marked with orange cross marks. Note that up to this point, no new actual samples have been generated. 
In addition, one identifies \textit{candidate strata} for refining the SS design by dividing the existing strata, which is shown schematically in Fig. \ref{fig:illustration of RLSS (b)} with blue solid lines.

The third step involves generating new \textit{candidate samples} for RLSS. 
In this sense, candidate samples are those that may \textcolor{black}{include} the already existing $\mathcal{N}$ samples. 
These candidate samples must be identified following a special procedure such that the properties of LSS continue being fulfilled. 
For materializing this third step, one must identify the strata which must contain candidate samples in order to enforce the LSS condition, and the strata where candidate samples can be generated randomly. 
This is illustrated schematically in Fig. \ref{fig:illustration of RLSS (c)}. 
The pink color indicates those strata that must contain candidate samples, while the yellow color shows those strata where a candidate sample may be generated at random. 
With all these considerations, one can generate $\mathcal{N}\delta$ candidate designs, as shown schematically in Fig. \ref{fig:illustration of RLSS (c)} with 4 orange dots.

The fourth step of RLSS is to incorporate a batch of $\hbar$ samples to the existing set of $\mathcal{N}$ samples. 
This is performed by selecting at random from the existing $\mathcal{N}\delta$ candidate samples. 
Note that in this process, it is necessary to update the strata associated with SS taking into account the candidate strata defined in the second step. 
Clearly, in such update, one must also update the weights (areas) of the selected strata. 
Fig. \ref{fig:illustration of RLSS (d)} illustrates the case where $\hbar=4$ and also shows the updated strata with green solid line.

It should be mentioned that the fourth step described above can be repeated as many times as necessary to select many batches of $\hbar$ samples as long as there are candidate samples left. 
In case one runs out of candidate samples, it is necessary to return to the second step and perform a new run of HLHS, which implies subdividing the strata associated with LHS. 
Furthermore, after each sample size extension, generated RLSS samples contain not only batches of additional samples, but also samples from the initial LSS design. 
In this work, we take $\hbar\ge\mathcal{N}$ \textcolor{black}{in order to include} the initial LSS design in the initial \textcolor{black}{RLSS} samples when $l=1$ in Eq. (\ref{eq:raw_moments_RLSS}). 

In the proposed sampling scheme, a proper convergence criterion should be developed to determine the desired number of sample size extensions. 
It is found that higher-order fractional moment always exhibits larger variability than its lower-order counterpart. 
Accordingly, if the variability of maximum order fractional moment is controlled, the variability of the lower-order ones will be automatically below a desired level.
Note that the maximum order of fractional moments is set to be 2 in this work, as mentioned in Section \textcolor{black}{\ref{section:concept_FM}}.
Therefore, a convergence criterion is proposed by judging the variability of the second-order fractional moment $\hat{M}_\mathcal{Z}^2$ evaluated by RLSS.
Specifically, the coefficient of variation (COV) of the $\hat{M}_\mathcal{Z}^2$ is compared with a user-defined small value $\mathcal{E}$ (e.g., $\mathcal{E} = 0.02$) to determine when to stop the sample size extension. 
The stopping criterion is defined as:
\begin{equation}\label{eq:convergence_criterion}
	\mathrm{COV} \left\{ \hat{M}_\mathcal{Z}^2 \right\} < \mathcal{E}.
\end{equation}
Although the expression of $\mathrm{COV} \left\{ \hat{M}_\mathcal{Z}^2 \right\}$ is not available for RLSS, the bootstrap resampling technique \cite{efron1994introduction} can be alternatively implemented here to estimate it.
Note that traditional bootstrap method generates samples with equal probability of occurrence, which \textcolor{black}{is not the case for} RLSS samples. 
To consider the unequal weight property of RLSS samples, the approach proposed in \textcolor{black}{Ref.} \cite{shields2015refined} is adopted here, such that samples with higher weights have more probability of being chosen for bootstrap. 
For more details on this approach, it is referred to \textcolor{black}{Ref.} \cite{shields2015refined}.

With such parallel adaptive scheme above, once the samples of $\mathcal{Z}$ that meet the convergence condition are obtained, a set of lower-order (only up to 2) fractional moments can be estimated according to Eq. (\ref{eq:raw_moments_RLSS}), which are then used to represent the EVD. 
	
\subsection{Representing EVD by \textcolor{black}{a mixture distribution} with fractional moments}
After obtaining the fractional moments of $\mathcal{Z}$, an adequate probability distribution model should be employed for the EVD estimation. 
Generally, the state-of-art distribution models represent the EVD by adopting either maximum entropy density \cite{xu2019novel,chen2020efficient} or positively skewed distributions such as shifted generalized lognormal distribution \cite{chen2021seismic} and a mixture of lognormal distribution and inverse Gaussian distribution \cite{dang2020mixture}. 
However, their flexibility is still limited for the EVDs with heavy tails, leading to \textcolor{black}{the inaccuracy of EVD reconstruction for some first-passage problems}. 
To increase the flexibility and enlarge the application scope, we first extend the traditional inverse Gaussian distributions by introducing an exponential transformation with an additional shape parameter.
Then, we introduce the log transformation to the extended skew-normal distribution, to enhance its ability to accommodate fat tails.
Further, these two improved distributions are mixed together to produce a more flexible mixture distribution model, whose involved parameters can be estimated from the estimated fractional moments.

\subsubsection{Proposed extended inverse Gaussian distribution}
The inverse Gaussian distribution (IGD) is a two-parameter skewed unimodal distribution and applies for positive real values \cite{folks1978inverse}. 
It is a first-passage time distribution for the Brownian motion with positive drift \cite{chhikara1988inverse}.
The PDF of \textcolor{black}{the} IGD is:
	\begin{equation}
	f_{\mathrm{IGD}}\left( z; a, b \right) =\sqrt{\frac{b}{2\pi z^3}}\exp \left[ -\frac{b\left( z-a \right) ^2}{2za^2} \right] , \;\; \mathrm{with} \; z>0,
	\end{equation}
where $a > 0$ is the location parameter; $b>0$ is the shape parameter. 

Denote the random variable which follows an IGD as $\mathcal{Z}_\mathrm{IGD}$. 
The $r$-th fractional moment of $\mathcal{Z}_\mathrm{IGD}$ is given as:    	
 \begin{equation}\label{eq:moment_inverseGaussian}
 	{M}_{\mathcal{Z}_{\mathrm{IGD}}}^r =
	E\left[ \mathcal{Z}_{\mathrm{IGD}}^{r} \right] =\int_0^{+\infty}{z^rf_{\mathrm{IGD}}\left( z \right)}\mathrm{d}z=\exp \left[ \frac{b}{a} \right] \sqrt{\frac{2b}{\pi}}a^{r-1/2}K_{1/2-r}\left( \frac{b}{a} \right), 
	\end{equation}
where $K_\alpha(\beta)$ is the modified Bessel function of the second kind.

In fact, the IGD can be extended to obtain higher flexibility in its shape. 
Here, we introduce a transformation $X=\mathcal{Z}^{1/\eta}$ to extend the original distribution, where $\eta > 0$ is a shape parameter. 
The resulting distribution is called extended inverse Gaussian distribution (EIGD). 
To obtain the PDF and fractional moments of the EIGD, the following theorem is \textcolor{black}{first given}: 

\textbf{Theorem 1.} \textit{Assume $X$ and $\mathcal{Z}$ are two continuous and positive real-valued random variables, and $f_{\mathcal{Z}}\left( z \right)$ is already available. 
Let $X=\mathcal{Z}^{1/\eta}$ where $\eta > 0$ , then we have $f_X\left( x \right) = f_{\mathcal{Z}}\left( x^\eta \right) \cdot \eta \cdot x^{\eta -1}$. 
\textcolor{black}{Additionally, the $r$-th fractional moment of $X$ is $E\left[ X^{r} \right] = E\left[ {\mathcal{Z}}^{r/\eta} \right]$.}}

\textbf{Proof.} \textit{Since $X=\mathcal{Z}^{1/\eta}$, according to the principle of conservation of probability, it is straightforward to derive $f_{\mathcal{Z}}\left( z \right) \mathrm{d}z=f_X\left( x \right) dx$. 
Thus, the PDF of $X$ can be derived as $f_X\left( x \right) =f_{\mathcal{Z}}\left( z \right) \frac{\mathrm{d}z}{dx}=f_{\mathcal{Z}}\left( x^\eta \right) \cdot \eta \cdot x^{\eta -1}$. 
\textcolor{black}{We may also derive the relationship between the $r$-th fractional moment of $X$ and that of $\mathcal{Z}$ as $E\left[ X^r \right] =E\left[ \left( \mathcal{Z} ^{1/\eta} \right) ^r \right] =E\left[ \mathcal{Z} ^{r/\eta} \right] $.}}

Therefore, the PDF of EIGD reads:	
	\begin{equation}\label{eq:EIG_PDF}
	f_{\mathrm{EIGD}}\left( x; \eta , a, b \right) =\eta \sqrt{\frac{b}{2\pi}}x^{-\eta /2-1}\exp \left[ -\frac{b\left( x^{\eta}-a \right) ^2}{2x^{\eta}a^2} \right] , \;\; \mathrm{with} \; x>0.
	\end{equation}
	
Denote the random variable which follows the EIGD as ${X}_\mathrm{EIGD}$. 
According to Eq. (\ref{eq:moment_inverseGaussian}) and \textbf{Theorem 1}, the $r$-th fractional moment of $X_\mathrm{EIGD}$ can be derived in \textcolor{black}{analytic} form:	
	\begin{equation}\label{eq:EIG_moment}
	{M}_{X_{\mathrm{EIGD}}}^r = \exp \left[ \frac{b}{a} \right] \sqrt{\frac{2b}{\pi}}a^{r/\eta -1/2}K_{1/2-r/\eta}\left( \frac{b}{a} \right). 
	\end{equation}
	
\begin{figure}[htb]
	\centering
	\includegraphics[width=0.5\linewidth]{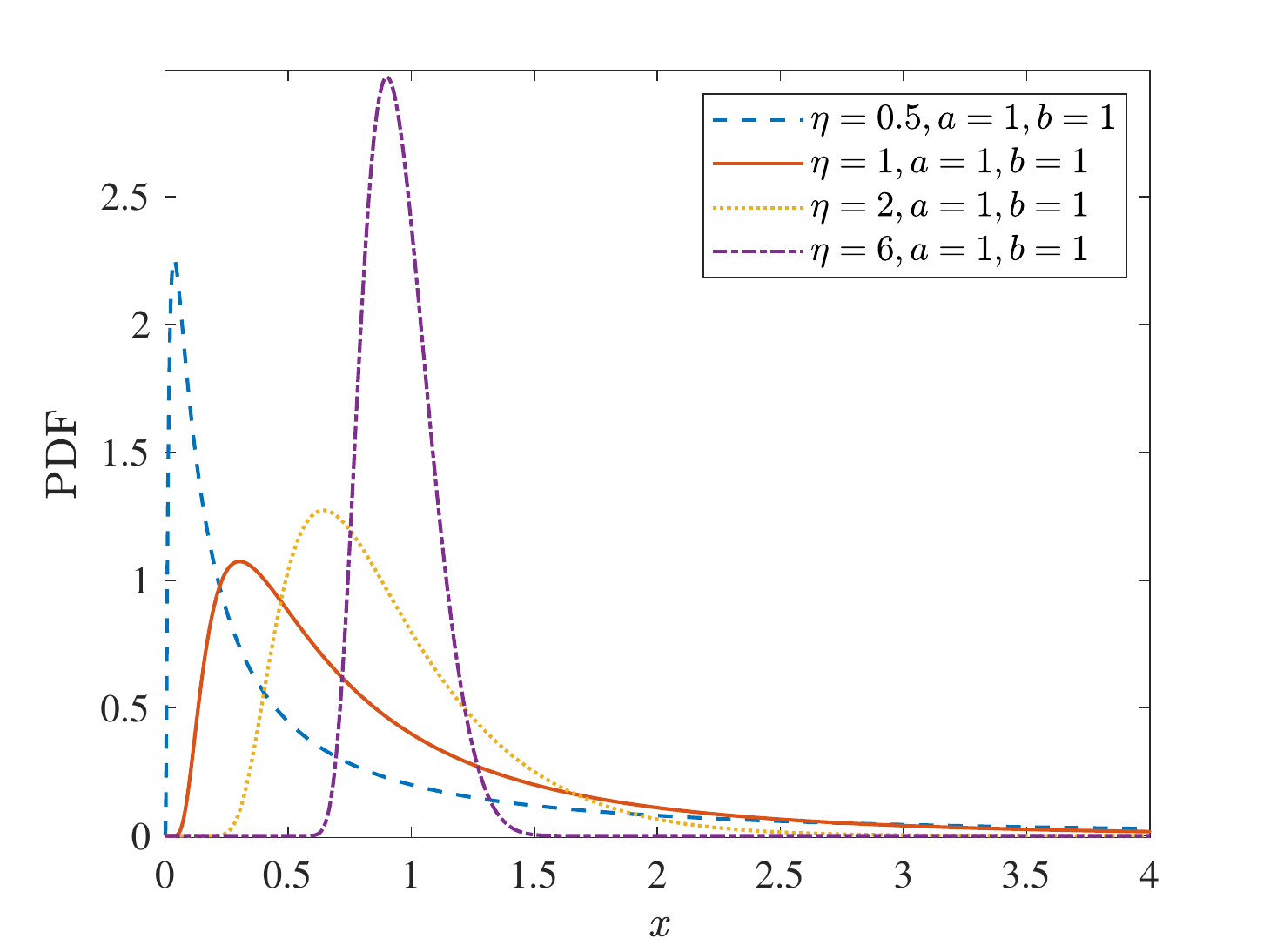}
	\caption{PDFs of extended inverse Gaussian distribution under four different sets of parameters}
	\label{fig:EIGflexibility}	
\end{figure}

Note that when $\eta = 1$, the EIGD reduces to the IGD according to Eq. (\ref{eq:EIG_PDF}).
The limit or special cases of IGD also belong to the EIGD, such as the chi-square distribution with single degree of freedom, normal distribution and Lévy distribution. 
Besides, the shape flexibility of \textcolor{black}{the} EIGD is illustrated by Fig. \ref{fig:EIGflexibility} under four different sets of parameters. 
In this figure, we make a comparison between the original IGD and the proposed EIGD by changing parameter $\eta$ and fixing $a=1,b=1$ of the EIGD.
It can be observed that, the proposed EIGD possesses much more flexibility in shape of PDF than the original IGD.

\subsubsection{Proposed log extended skew-normal distribution}
The extended skew-normal distribution (ESND) was first introduced by Azzalini \cite{azzalini2013skew}. 
This distribution is a four-parameter unimodal asymmetric distribution with support on $\left( -\infty, +\infty \right)$, which generalizes the traditional skew-normal distribution and normal distribution.
The statistical properties of the ESND are discussed in detail in Ref. \cite{canale2011statistical}.
The PDF of \textcolor{black}{the} ESND of a real random variable $\tilde{X} \in \mathbb{R}$ is:
\begin{equation}\label{eq:ESN_PDF}
    f_{\mathrm{ESND}}\left( \tilde{x};c,d,\theta ,\tau \right) =\frac{1}{d}\phi \left( \frac{\tilde{x}-c}{d} \right) \frac{\Phi \left( \tau \sqrt{1+\theta ^2}+\theta \frac{\tilde{x}-c}{d} \right)}{\Phi \left( \tau \right)},\;\;\mathrm{with}\;\tilde{x}\in \mathbb{R},
\end{equation}
where $c \in \mathbb{R}$ is the location parameter; $d>0$ is the scale parameter; $\theta \in \mathbb{R}$ is the shape parameter; $\tau \in \mathbb{R}$ is the truncation parameter; $\phi \left( \cdot \right)$ and  $\Phi \left( \cdot \right)$ denote the PDF and CDF of \textcolor{black}{the} standard normal distribution.

The moment-generating function (MGF) of \textcolor{black}{the} ESND is:
\begin{equation}\label{eq:MGF_ESN}
    M_{\tilde{X}}\left( \tilde{t} \right) =E\left[ \exp \left( \tilde{t}\tilde{X} \right) \right] =\exp \left( c\tilde{t}+\frac{1}{2}d^2\tilde{t}^2 \right) \frac{\Phi \left( \tau +\frac{\theta d\tilde{t}}{\sqrt{1+\theta ^2}} \right)}{\Phi \left( \tau \right)}, \;\;\mathrm{with}\;\tilde{t} \in \mathbb{R}.
\end{equation}

Although the ESND enables to accommodate asymmetry characteristics, its ability to fit heavier tails can be further improved by introducing \textcolor{black}{a} log transformation to the ESND. 
We shall refer the newly generated distribution as log extended skew-normal distribution (LESND).
Denote the random variable which follows a LESND as ${X}_\mathrm{LESND}$.
Then, we have the relationship between ${X}_\mathrm{LESND}$ and $\tilde{X}$ as  $X_{\mathrm{LESND}}=\exp \left( \tilde{X} \right)$.
That is, the logarithm of ${X}_\mathrm{LESND}$ follows the original ESND.
Hence, we can get the PDF of \textcolor{black}{the} LESND as:
	\begin{equation}\label{eq:LESN_PDF}
          f_{\mathrm{LESND}}\left( x;c,d,\theta ,\tau \right) =\frac{1}{dx}\phi \left( \frac{\log \left( x \right) -c}{d} \right) \frac{\Phi \left( \tau \sqrt{1+\theta ^2}+\theta \frac{\log \left( x \right) -c}{d} \right)}{\Phi \left( \tau \right)},\;\;\mathrm{with}\;x>0.
	\end{equation}

From the relationship between the fractional moment of \textcolor{black}{the} LESND and the MGF of \textcolor{black}{the} ESND, it is easy to derive ${M}_{X_{\mathrm{LESND}}}^{r}=E\left[ X_{\mathrm{LESND}}^{r} \right] =E\left[ \left( \exp \left( \tilde{X} \right) \right) ^r \right] = M_{\tilde{X}}\left( r \right)$. 
Hence, the $r$-th fractional moment of  ${X}_{\mathrm{LESND}}$ can be given in analytic form as:
 \begin{equation}\label{eq:moment_LESN}
 {M}_{X_{\mathrm{LESND}}}^{r}
 =\exp \left( cr+\frac{1}{2}d^2r^2 \right) \frac{\Phi \left( \tau +\frac{\theta dr}{\sqrt{1+\theta ^2}} \right)}{\Phi \left( \tau \right)}.
	\end{equation}

\begin{figure}[htb]
	\centering
	\includegraphics[width=0.5\linewidth]{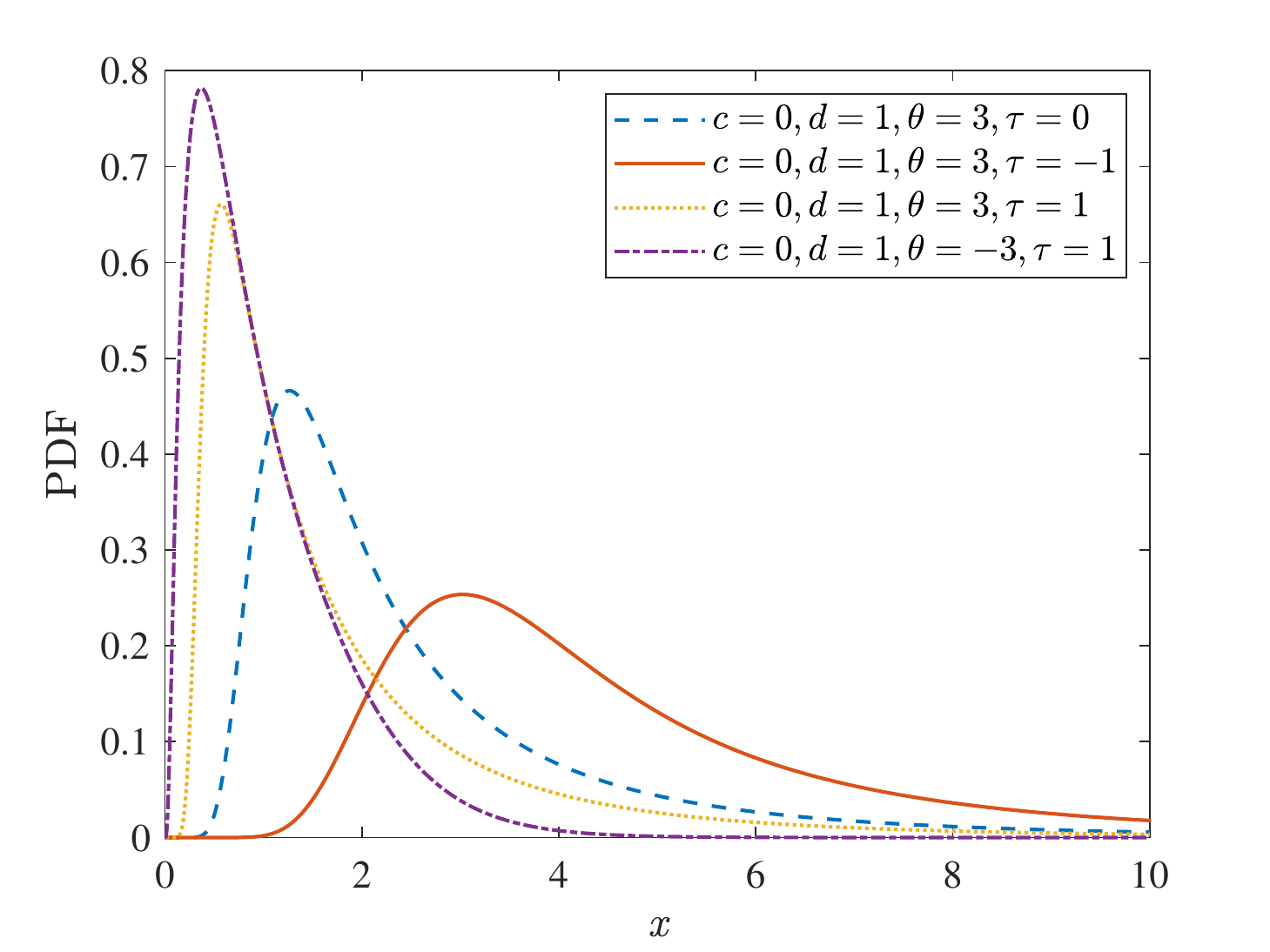}
	\caption{PDFs of log extended skew-normal distribution under four different sets of parameters}
	\label{fig:LESNflexibility}	
\end{figure}

Note that according to Eq. (\ref{eq:LESN_PDF}), when $\tau = 0$, the LESND reduces to the log skew-normal distribution \cite{wang2019novel}; and when $\theta = 0$, the LESND reduces to the traditional lognormal distribution. 
It should be mentioned that if $\theta = 0$, the shape of LESND will not be affected by changing the value of parameter $\tau$.
Besides, to illustrate the flexibility of the LESND, Fig. \ref{fig:LESNflexibility} depicts the LESND with four sets of parameters.  
In this figure, the log skew-normal distribution is given for comparison by setting the parameters of LESND as $c=0,d=1,\theta=3,\tau=0$. 
\textcolor{black}{As can be seen, the LESND provides richer distribution shapes compared to the log skew-normal distribution,} showing the increased flexibility of LESND. 

\subsubsection{Proposed mixture distribution}
It is worth mentioning that the first-passage probability estimation is closely associated to the distribution tail of EVD.
Besides, the EVD is usually asymmetric and possesses heavy tail in many cases.
Hence, a highly flexible distribution model is needed, which is suitable for fitting distributions with various tail properties, especially the heavy-tailed distributions. 
For accurate EVD estimation, two single-component skewed distributions proposed above, i.e., the EIGD and LESND, may still not be flexible enough \textcolor{black}{and their applicability to various first-passage problems is limited}. 
To further improve the flexibility, one potential way is to mix the proposed single-component distributions together by introducing a weight parameter. 
Such distribution model enables to incorporate both characteristics of two single-component distributions, and can accommodate asymmetry in a more flexible way so as to properly estimate the EVD. 
Therefore, motivated by the above, 
a novel mixture of the extended inverse Gaussian and log extended skew-normal distributions (\textcolor{black}{M-EIGD-LESND}) is developed herein.

The PDF of \textcolor{black}{M-EIGD-LESND} is given as:
\begin{equation}\label{eq:M-EIGD-LESND_PDF}
\begin{array}{l}
	f_{\mathrm{\textcolor{black}{M-EIGD-LESND}}}\left( x;\varUpsilon \right) =wf_{\mathrm{EIGD}}\left( x;\eta ,a,b \right) +\left( 1-w \right) f_{\mathrm{LESND}}\left( x;c,d,\theta ,\tau \right)\\
	=w\eta \sqrt{\frac{b}{2\pi}}x^{-\eta /2-1}\exp \left[ -\frac{b\left( x^{\eta}-a \right) ^2}{2x^{\eta}a^2} \right]
	+\left( 1-w \right) \frac{1}{dx}\phi \left( \frac{\log \left( x \right) -c}{d} \right) \frac{\Phi \left( \tau \sqrt{1+\theta ^2}+\theta \frac{\log \left( x \right) -c}{d} \right)}{\Phi \left( \tau \right)},\;\;\mathrm{with}\;x>0,\\
\end{array}
\end{equation}
where $\varUpsilon =\left[ w, \eta, a, b, c,d,\theta, \tau \right]$ is the set of eight unknown parameters and $w \in \left[0, 1\right]$ is the weight parameter of \textcolor{black}{M-EIGD-LESND}. 

According to Eqs. (\ref{eq:EIG_moment}) and (\ref{eq:moment_LESN}), the $r$-th fractional moment of \textcolor{black}{M-EIGD-LESND} can be given in \textcolor{black}{analytic} form:
\begin{equation}\label{eq:M-EIG-LESN_moment}
\begin{array}{l}
	{M}_{X_{\mathrm{\textcolor{black}{M-EIGD-LESND}}}}^{r}=E\left[ X_{\mathrm{\textcolor{black}{M-EIGD-LESND}}}^{r};\varUpsilon \right] =wE\left[ X_{\mathrm{EIGD}}^{r} \right] +\left( 1-w \right) E\left[ X_{\mathrm{LESND}}^{r} \right]\\
	=w\exp \left[ \frac{b}{a} \right] \sqrt{\frac{2b}{\pi}}a^{r/\eta -1/2}K_{1/2-r/\eta}\left( \frac{b}{a} \right) +\left( 1-w \right) \exp \left( cr+\frac{1}{2}d^2r^2 \right) \frac{\Phi \left( \tau +\frac{\theta dr}{\sqrt{1+\theta ^2}} \right)}{\Phi \left( \tau \right)}.\\
\end{array}
\end{equation}

\begin{figure}[htb]
	\centering
	\includegraphics[width=0.6\linewidth]{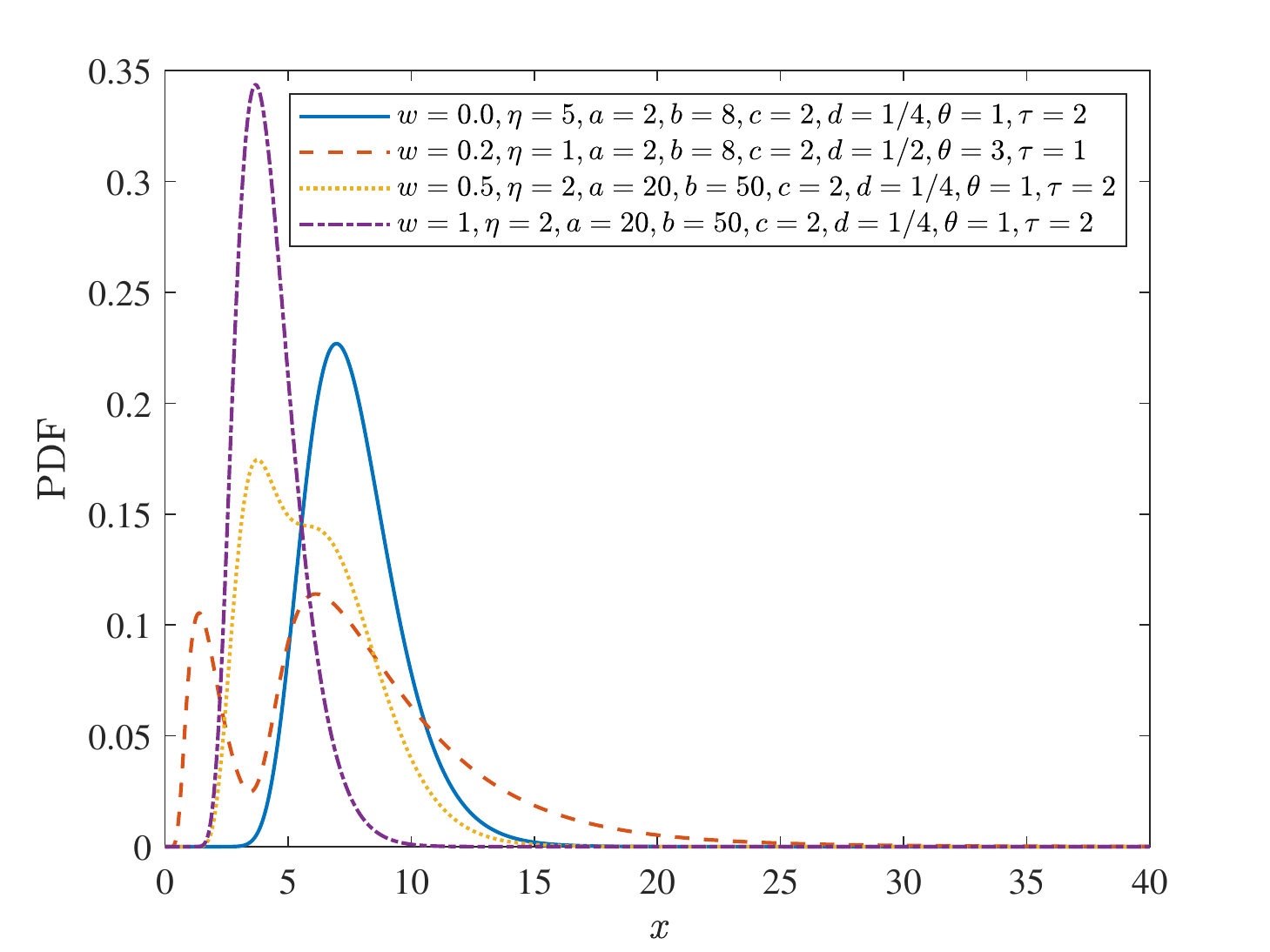}
	\caption{\textcolor{black}{PDFs of the proposed mixture distribution under four different sets of parameters}}
	\label{fig:Mixture_dist_shape}	
\end{figure}

Note that the proposed \textcolor{black}{M-EIGD-LESND} can \textcolor{black}{reduce} to the mixture of lognormal and inverse Gaussian distributions \cite{dang2020mixture} if set $\eta = 1$ and $\theta = 0$. 
To illustrate the flexibility of the proposed mixture distribution model, Fig. \ref{fig:Mixture_dist_shape} shows the plot of the PDFs associated with \textcolor{black}{M-EIGD-LESND} with different parameters.
It can be seen that the proposed mixture distribution model is highly flexible with rich shapes \textcolor{black}{and enables to accommodate various heavy tails. 
In addition, the \textcolor{black}{M-EIGD-LESND} is able to represent} not only unimodal distributions but also bimodal distributions. 

\subsubsection{Parameter estimation}\label{section:Parameter_estimation}
The proposed mixture distribution model has the potential to characterize the EVD. 
\textcolor{black}{Hence, in order to recover the EVD of $\mathcal{Z}$}, we assume that the EVD follows the proposed mixture distribution model, and determine the free parameters of this model in an appropriate way. 
Note that the proposed distribution contains a set of eight free parameters. 
To estimate these unknown distribution parameters, a natural way is to match the fractional moments of \textcolor{black}{the proposed mixture distribution model with the estimated fractional moments of the corresponding orders} (hereafter referred to as the fractional moment matching technique). 
Accordingly, the following nonlinear system of equations requires to be solved: 
\begin{equation}\label{eq:Nonlin_equations_para}
\left\{ \begin{array}{c}
	\hat{M}_{\mathcal{Z}}^{r_1}={M}_{X_{\mathrm{\textcolor{black}{M-EIGD-LESND}}}}^{r_1}\\
	\hat{M}_{\mathcal{Z}}^{r_2}={M}_{X_{\mathrm{\textcolor{black}{M-EIGD-LESND}}}}^{r_2}\\
	\cdots\\
	\hat{M}_{\mathcal{Z}}^{r_8}={M}_{X_{\mathrm{\textcolor{black}{M-EIGD-LESND}}}}^{r_8},\\
\end{array} \right. 
\end{equation}
where $\hat{M}_{\mathcal{Z}}^{r_i}, i=1,2,...,8$ are the ${r_i}$-th fractional moments estimated by RLSS; 
${M}_{X_{\mathrm{\textcolor{black}{M-EIGD-LESND}}}}^{r_i}$ can be \textcolor{black}{obtained} by Eq. (\ref{eq:M-EIG-LESN_moment}); 
and the fractional order $r_i$ takes $\left[{r_1},{r_2},...,{r_8} \right] = \frac{2}{8} \times \left[1,2,...,8\right]$. 
\textcolor{black}{Here, the equally spaced fractional orders are introduced for convenience, since it is straightforward to take such value without any prior knowledge of fractional orders. 
Besides, as adopted in Ref. \cite{dang2020mixture}, the maximum fractional order is set to be 2, since the second-order fractional moment can be estimated efficiently from a small number of samples, and it reflects more shape information of EVD than lower-order fractional moments, as discussed in Section \ref{section:concept_FM}. }

Solution to Eq. (\ref{eq:Nonlin_equations_para}) can be obtained \textcolor{black}{in seconds} by any appropriate nonlinear solver, such as \textit{lsqnonlin} in Matlab. 
To facilitate the solving process, initial values for the free parameters are required. 
Denote the initial values of Eq. (\ref{eq:Nonlin_equations_para}) as $ w_{0}, \hat{\eta} _0,\hat{a}_0,\hat{b}_0, \hat{c}_0, \hat{d}_0, \hat{\theta}_0, \hat{\tau}_0$. 
$w_{0}$ is set to be 0.5 to \textcolor{black}{assign an equal initial weights to the two single-component functions.} 
The other initial values, i.e., $ \hat{\eta} _0,\hat{a}_0,\hat{b}_0, \hat{c}_0, \hat{d}_0, \hat{\theta}_0, \hat{\tau}_0$, can be obtained by another low-order fractional moment matching technique, where a nonlinear system of equations is involved: 
\begin{equation}\label{eq:initial_optim1}
\left\{ \begin{array}{c}
	\hat{M}_{\mathcal{Z}}^{1/2}=M_{X_{\mathrm{EIGD}}}^{1/2}\\
	\hat{M}_{\mathcal{Z}}^{1}=M_{X_{\mathrm{EIGD}}}^{1}\\
	\hat{M}_{\mathcal{Z}}^{3/2}=M_{X_{\mathrm{EIGD}}}^{3/2}\\
\end{array}, \right.
\end{equation} 
and 
\begin{equation}\label{eq:initial_optim2}
\left\{ \begin{array}{c}
	\hat{M}_{\mathcal{Z}}^{1/2}=M_{X_{\mathrm{LESND}}}^{1/2}\\
	\hat{M}_{\mathcal{Z}}^{1}=M_{X_{\mathrm{LESND}}}^{1}\\
	\hat{M}_{\mathcal{Z}}^{3/2}=M_{X_{\mathrm{LESND}}}^{3/2}\\
	\hat{M}_{\mathcal{Z}}^{2}=M_{X_{\mathrm{LESND}}}^{2}\\
\end{array}, \right. 
\end{equation}  
where $\hat{\eta}_0 >0,\; \hat{a}_0>0, \; \hat{b}_0>0, \hat{c}_0 \in \mathbb{R}, \; \hat{d}_0> 0, \; \hat{\theta}_0 \in \mathbb{R},\; \hat{\tau}_0 \in \mathbb{R}$.
Note that the \textcolor{black}{M-EIGD-LESND} can reduce to \textcolor{black}{the} inverse Gaussian distribution (if set $w=0, \eta = 1$) or \textcolor{black}{the} lognormal distribution (if set $w=1, \theta = 0$), and the relationships between the parameters and the first two central moments of each reduced distribution are easy to be obtained. 
Besides, as discussed earlier, the value of parameter $\tau$ will be irrelevant if $\theta = 0$.
Hence, the initial values for Eqs (\ref{eq:initial_optim1}) and (\ref{eq:initial_optim2}) can be determined as: $a_0=\hat{\mu}_{\mathcal{Z}},b_0=\hat{\mu}_{\mathcal{Z}}^{3}/\hat{\sigma}_{\mathcal{Z}}^{2},\eta _0=1
$, $c_0=\log \left( \hat{\mu}_{\mathcal{Z}}^{2}/\sqrt{\hat{\sigma}_{\mathcal{Z}}^{2}+\hat{\mu}_{\mathcal{Z}}^{2}} \right) ,d_0=\sqrt{\log \left( \hat{\sigma}_{\mathcal{Z}}^{2}/\hat{\mu}_{\mathcal{Z}}^{2}+1 \right)},\theta _0=0,\tau _0=0
$, where $\hat{\mu} _{\mathcal{Z}} = \hat{M}_{\mathcal{Z}}^{1}$ and $\hat{\sigma} _{\mathcal{Z}} = \sqrt{\hat{M}_{\mathcal{Z}}^{2} - \left(\hat{M}_{\mathcal{Z}}^{1}\right)^2}$.
The parameter estimation process of proposed \textcolor{black}{M-EIGD-LESND} is briefly summarized in Algorithm \ref{alg:Algorithm_2}.

\begin{algorithm}[htb!]
	\caption{Parameter estimation for \textcolor{black}{M-EIGD-LESND} using the fractional moment matching technique} 
	\label{alg:Algorithm_2}
	\hspace*{0.02in} {\bf Input:} 
	central moments $\hat{\mu}_{\mathcal{Z}}$, $\hat{\sigma}_{\mathcal{Z}}$, and fractional moments $\hat{M}_{\mathcal{Z}}^{\boldsymbol{r}}$ ($\boldsymbol{r} = \left[\frac{1}{4}, \frac{1}{2}, \frac{3}{4}, 1, \frac{5}{4}, \frac{3}{2}, \frac{7}{4}, 2\right]$). 
	
	\hspace*{0.02in} {\bf Output:} 
	estimated distribution parameters $\varUpsilon = \left[ w, \eta, a, b, c, d, \theta, \tau \right]$.
	\begin{algorithmic}[1]
		\State Use $\hat{\mu}_{\mathcal{Z}}$ and $\hat{\sigma}_{\mathcal{Z}}$ to evaluate $\eta _0,a_0,b_0,c_0,d_0,\theta _0,\tau _0$ as the initial values of Eqs. (\ref{eq:initial_optim1}) and (\ref{eq:initial_optim2});
        \State Solve Eqs. (\ref{eq:initial_optim1}) and (\ref{eq:initial_optim2}) with $\eta _0,a_0,b_0,c_0,d_0,\theta _0,\tau _0$ to estimate the initial values $\hat{\eta} _0,\hat{a}_0,\hat{b}_0, \hat{c}_0, \hat{d}_0, \hat{\theta}_0, \hat{\tau}_0$ of Eq. (\ref{eq:Nonlin_equations_para}). 
        \State Solve the fractional moment matching equations (Eq. (\ref{eq:Nonlin_equations_para})) by means of any appropriate nonlinear solver with $\hat{\eta} _0,\hat{a}_0,\hat{b}_0, \hat{c}_0, \hat{d}_0, \hat{\theta}_0, \hat{\tau}_0$ and $w_{0}=0.5$, and then obtain the estimated distribution parameters $\varUpsilon = \left[ w, \eta, a, b, c, d, \theta, \tau \right]$ of \textcolor{black}{M-EIGD-LESND}.
	\end{algorithmic}
\end{algorithm}



\subsection{Procedure of the proposed method}
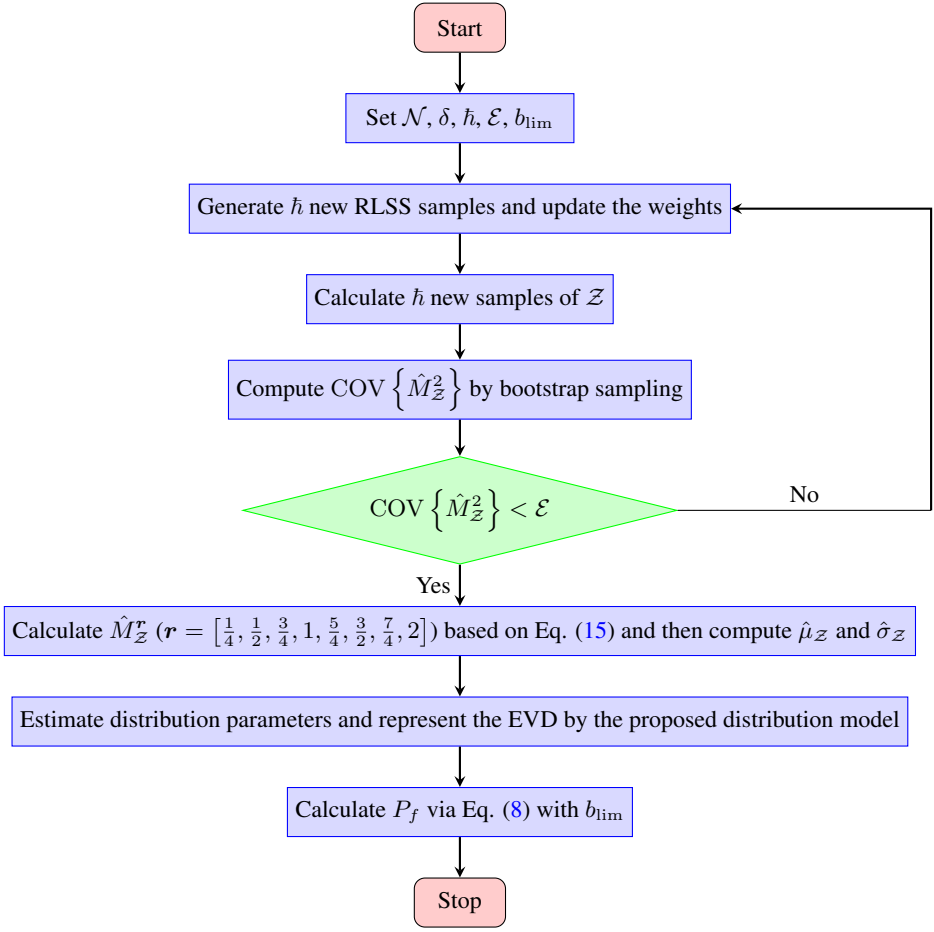
\begin{figure}[!htb]
    \centering
	\tikzstyle{startstop} = [rectangle,rounded corners, minimum width=1.2cm, minimum height=0.65cm, text centered, draw=black, fill=red!20]
	\tikzstyle{io} = [trapezium, trapezium left angle = 70, trapezium right angle=110, minimum width=3cm, minimum height=1cm, text centered, draw=black, fill=blue!20]
	\tikzstyle{process} = [rectangle, minimum width=3cm, minimum height=0.65cm, text centered,  draw=blue, fill=blue!15]
	\tikzstyle{decision} = [diamond,minimum width=1cm,minimum height=0.05cm,text centered,draw=green,fill=green!20,aspect=4]
	\tikzstyle{arrow} = [thick,->,>=stealth]
	\small
	
	\begin{tikzpicture}[node distance=1.2cm]
	\node (start) [startstop] {\textcolor{black}{Start}};
	\node (process1) [process,below of=start] {Set $\mathcal{N}$, $\delta$, $\hbar$, $\mathcal{E}$, $b_{\mathrm{lim}}$};
	\node (process2) [process,below of=process1] {Generate $\hbar$ new RLSS samples and update the weights};
	\node (process3) [process,below of=process2] {Calculate $\hbar$ new samples of $\mathcal{Z}$};
	\node (process4) [process,below of=process3] {Compute $\mathrm{COV}\left\{ \hat{M}_\mathcal{Z}^2 \right\}$ by bootstrap sampling};
	\node (decision1) [decision,below of=process4,yshift=-0.4cm] {$\mathrm{COV}\left\{ \hat{M}_\mathcal{Z}^2 \right\}<\mathcal{E}$};
	\node (process5) [process,below of=decision1,yshift=-0.4cm] {Calculate $\hat{M}_{\mathcal{Z}}^{\boldsymbol{r}}$ ($\boldsymbol{r} = \left[\frac{1}{4}, \frac{1}{2}, \frac{3}{4}, 1, \frac{5}{4}, \frac{3}{2}, \frac{7}{4}, 2\right]$) based on Eq. (\ref{eq:raw_moments_RLSS}) \textcolor{black}{and then compute $\hat{\mu}_\mathcal{Z}$ and $\hat{\sigma}_\mathcal{Z}$}};
	\node (process6) [process,below of = process5] {Estimate distribution parameters and represent the EVD by the proposed distribution model};
	\node (process7) [process,below of = process6] {Calculate $P_f$ via Eq. (\ref{eq:MEVD2}) with $b_{\mathrm{lim}}$};
	\node (stop) [startstop,below of=process7] {\textcolor{black}{Stop}};
	
	\node[shape=coordinate](y1) [right of = decision1, xshift= 5 cm] {Y1};
	
	\draw [arrow] (start) -- (process1);
	\draw [arrow] (process1) -- (process2);
	\draw [arrow] (process2) -- (process3);
	\draw [arrow] (process3) -- (process4);
	\draw [arrow] (process4) -- (decision1);
	\draw [arrow] (decision1) -- node[anchor=east] {Yes} (process5);
	\draw  (decision1) -- node[above] {No} (y1);
	\draw [arrow] (y1) |- (process2);
	\draw [arrow] (process5) -- (process6);
	\draw [arrow] (process6) -- (process7);
	\draw [arrow] (process7) -- (stop);
	\end{tikzpicture}
	\caption{Flowchart of the proposed method}
	\label{fig:flowchart}
\end{figure}

Once the EVD is reconstructed by the proposed probability distribution model, the first-passage probability can be evaluated by Eq. (\ref{eq:MEVD2}) \textcolor{black}{for} a given threshold. A flowchart of the proposed method is shown in Fig. \ref{fig:flowchart}, and a brief procedure is summarized as follows:
    ~\\
    
	\textbf{Step 1}: Initialization. Set the initial sample size $\mathcal{N}$ of LSS, the refinement factor $\delta$ of HLHS, the number of samples $\hbar$ added in each sample size extension and \textcolor{black}{the value of tolerance $\mathcal{E}$.
	Determine the threshold $b_{\mathrm{lim}}$.} 


	\textbf{Step 2}: Generate $\hbar$ new samples by RLSS. Produce $\hbar$ new samples and update the weights by RLSS method according to Algorithm \ref{alg:algorithm_1} in \ref{section:appendix}, and then compute the new samples of $\mathcal{Z}$. 
	
	\textbf{Step 3}: Judge the convergence criterion. Evaluate the COV of $\hat{M}_{\mathcal{Z}}^2$ by using bootstrap technique. If Eq. (\ref{eq:convergence_criterion}) is satisfied, then turn to step 4; otherwise, return to step 2. 
	
	\textbf{Step 4}: Moment evaluation. Calculate a set of fractional moments $\hat{M}_{\mathcal{Z}}^{\boldsymbol{r}}$ ($\boldsymbol{r} = \left[\frac{1}{4}, \frac{1}{2}, \frac{3}{4}, 1, \frac{5}{4}, \frac{3}{2}, \frac{7}{4}, 2\right]$) according to Eq. (\ref{eq:raw_moments_RLSS}), \textcolor{black}{and then compute the first-two central moments $\hat{\mu}_{\mathcal{Z}}$ and $\hat{\sigma}_{\mathcal{Z}}$ by $\hat{\mu}_{\mathcal{Z}} = \hat{M}_{\mathcal{Z}}^{1}$ and $\hat{\sigma}_{\mathcal{Z}} = \sqrt{\hat{M}_{\mathcal{Z}}^{2} - \left(\hat{M}_{\mathcal{Z}}^{1}\right)^2}$}.
	
	\textbf{Step 5}: EVD representation. Represent the EVD by using the proposed distribution model, i.e., \textcolor{black}{M-EIGD-LESND}, where the involved free distribution parameters are estimated by the low-order fractional moment matching technique described in Algorithm \ref{alg:Algorithm_2}. 
	
	\textbf{Step 6}: First-passage probability estimation. Evaluate the first-passage probability $P_f = \mathrm{Pr}\left\{ \mathcal{Z} >b_{\lim} \right\}$ via \textcolor{black}{obtained EVD and Eq. (\ref{eq:MEVD2}).}
    ~\\	
	
\section{Numerical examples}\label{section:section4}
In this section, three examples, including two test examples and one practical engineering example, will be investigated to verify the efficacy of the proposed method. 
In all examples, the parameters of the proposed method are set as $\mathcal{N}=1$, $\delta = 1$, $\hbar = 8$ and $\mathcal{E} = 0.015$.
The computational efficiency and accuracy of proposed methods for first-passage probability estimation are compared with MCS, Subset simulation (SS) \cite{au2001estimation,au2007application} and two state-of-art mixture distribution methods presented in Ref. \cite{dang2020mixture} and \cite{dang2021approach}.
Note that in SS, the number of samples per layer is 1000 and the conditional probability is 0.1. 
Both the existing \textcolor{black}{mixture distribution} methods for comparison employ the Latinized partially stratified sampling (LPSS) to evaluate fractional moments of $\mathcal{Z}$.  
\textcolor{black}{The mixture distribution method in Ref.} \cite{dang2020mixture} develops a mixture distribution combining conventional inverse Gaussian and lognormal distributions (MIGLD), and thus this method is referred as LPSS+MIGLD in the following examples.
Another existing \textcolor{black}{mixture distribution} method \cite{dang2021approach} develops a mixture of two generalized inverse Gaussian distributions (MTGIG), and this method is denoted as LPSS+MTGIG in the following examples.

\subsection{Example 1: a Duffing oscillator under Gaussian white noise}
The first example considers a Duffing oscillator with uncertain parameters under Gaussian white noise, which is governed by
	\begin{equation}\label{eq:EX1 equation of motion}
\ddot{Y}\left( t \right) +\gamma \dot{Y}\left( t \right) +Y\left( t \right) +\varepsilon Y^3\left( t \right) =\mathscr{G} \left( t \right), 
	\end{equation}
where $\ddot{Y}$, $\dot{Y}$ and $Y$ are the acceleration, velocity and displacement at time $t$; $\gamma$ denotes the damping control coefficient; $\varepsilon$ is the parameter controlling the degree of nonlinearity in the restoring force; and $\mathscr{G} \left( t \right) $ is the Gaussian white noise. 
\textcolor{black}{Differential equation} solver \textit{Ode45} in Matlab is utilized to solve Eq. (\ref{eq:EX1 equation of motion}). 
Both $\gamma$ and $\varepsilon$ follow the lognormal distributions with mean values as 0.5 and 0.3, and standard deviation values as 0.2 and 0.1, respectively. 
The Gaussian white noise is expressed as
\begin{equation}
\mathscr{G} \left( t_k \right) =\theta \left( t_k \right) \sqrt{2\pi S/\varDelta t},
\end{equation}
where $S = 1$ is the spectral intensity; $\varDelta t = 0.01 \;\mathrm{s}$ is the time interval; $T = 30 \;\mathrm{s}$ is the time period; $t_k = k \varDelta t, k = 0,1,...,n_t$ is the discrete time; and here we consider $n_t = T / \varDelta t + 1 = 3001$ random variables $\theta \left( t_k \right)$ in the Gaussian white noise following the standard normal distributions. 
Therefore, a total number of $2+n_t=3003$ random variables are involved in the present example.

The maximum absolute extreme value of displacement over time ${t\in \left[ 0,\left. T \right] \right.}$, i.e., $\mathcal{Z} =\max_{t\in \left[ 0,\left. T \right] \right.} \left\{ \left| Y\left( t \right) \right| \right\}$, is of interest in this example. 
First, the proposed parallel adaptive sampling scheme is implemented for fractional moment estimation. 
The proposed scheme performs sample size extension successively until the convergence criterion in Eq. (\ref{eq:convergence_criterion}) is satisfied.
In each sample size extension, $\hbar = 8$ new RLSS samples are firstly generated for deterministic dynamic analysis.
Then, 8 new samples of $\mathcal{Z}$ are produced at a time using parallel computing technique with 8 CPU processors.
After that, the RLSS weights are redistributed so that the weights produced by all performed sample size extensions sum to 1. 
Subsequently, Eq. (\ref{eq:convergence_criterion}) is 
checked to determine whether to perform a new round of sample size extension.
Accordingly, a total of $\hat{\mathcal{N}}=520$ samples of $\mathcal{Z}$ are produced that satisfy the convergence criterion, where the corresponding $\hat{M}_{\mathcal{Z}}^{\boldsymbol{r}}$ ($\boldsymbol{r} = \left[\frac{1}{4}, \frac{1}{2}, \frac{3}{4}, 1, \frac{5}{4}, \frac{3}{2}, \frac{7}{4}, 2\right]$) can be obtained by Eq. (\ref{eq:raw_moments_RLSS}). 
Table \ref{tab:mom_comparison_example1} compares the first-two central moments ($\hat{\mu}_{\mathcal{Z}} = \hat{M}_{\mathcal{Z}}^{1}$ and $\hat{\sigma}_{\mathcal{Z}} = \sqrt{\hat{M}_{\mathcal{Z}}^{2} - \left(\hat{M}_{\mathcal{Z}}^{1}\right)^2}$) with the benchmark results given by MCS with $10^6$ runs. 
In this table, relative errors of the first-two moments between proposed method and MCS are also given, i.e., 0.5656\% and 0.7195\%, which indicate that proposed parallel adaptive scheme using RLSS enables to obtain accurate low-order central moments.

\begin{table}[htb]
	\caption{Comparison of first-two central moments by the proposed method and MCS (Example 1)}
	\label{tab:mom_comparison_example1}
	\centering 	
	\begin{tabular}{lll}
		\toprule 
		Method($\hat{\mathcal{N}}$)	  &$\hat{\mu}_{\mathcal{Z}}$       & $\hat{\sigma}_{\mathcal{Z}}$     \\
		\midrule       
		Proposed(520)  &  3.6570   &  0.6623      \\
		MCS($10^6$)       & 3.6778    &  0.6671     \\
		R.E.       & 0.5656\%    &  0.7195\%     \\
		\bottomrule       
	\end{tabular}
\leftline{~~~~~~~~~~~~~~~~~~~~~~~~~~~~~~~~~~~~~~~~~~~~~~~~~~~~~~~~~~~~Note: R.E. = Relative error with reference to MCS.}
\end{table}

\begin{figure}[!htb]
	\centering
	\subfigure[PDF]{
		\begin{minipage}{8.0cm}
			\centering
			\includegraphics[scale=0.50]{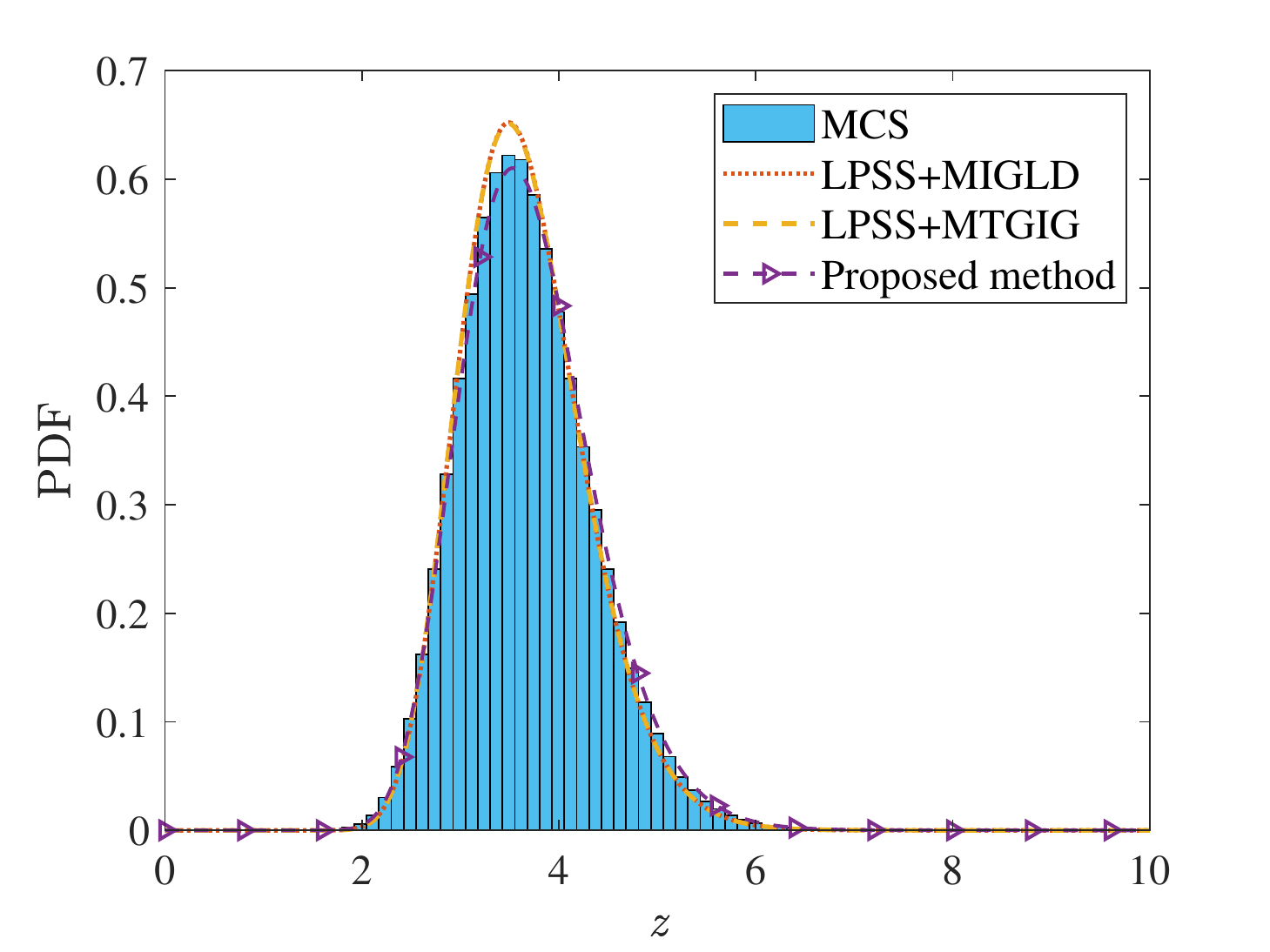}
		\end{minipage}
	}%
	\subfigure[CDF]{
		\begin{minipage}{8.0cm}
			\centering
			\includegraphics[scale=0.50]{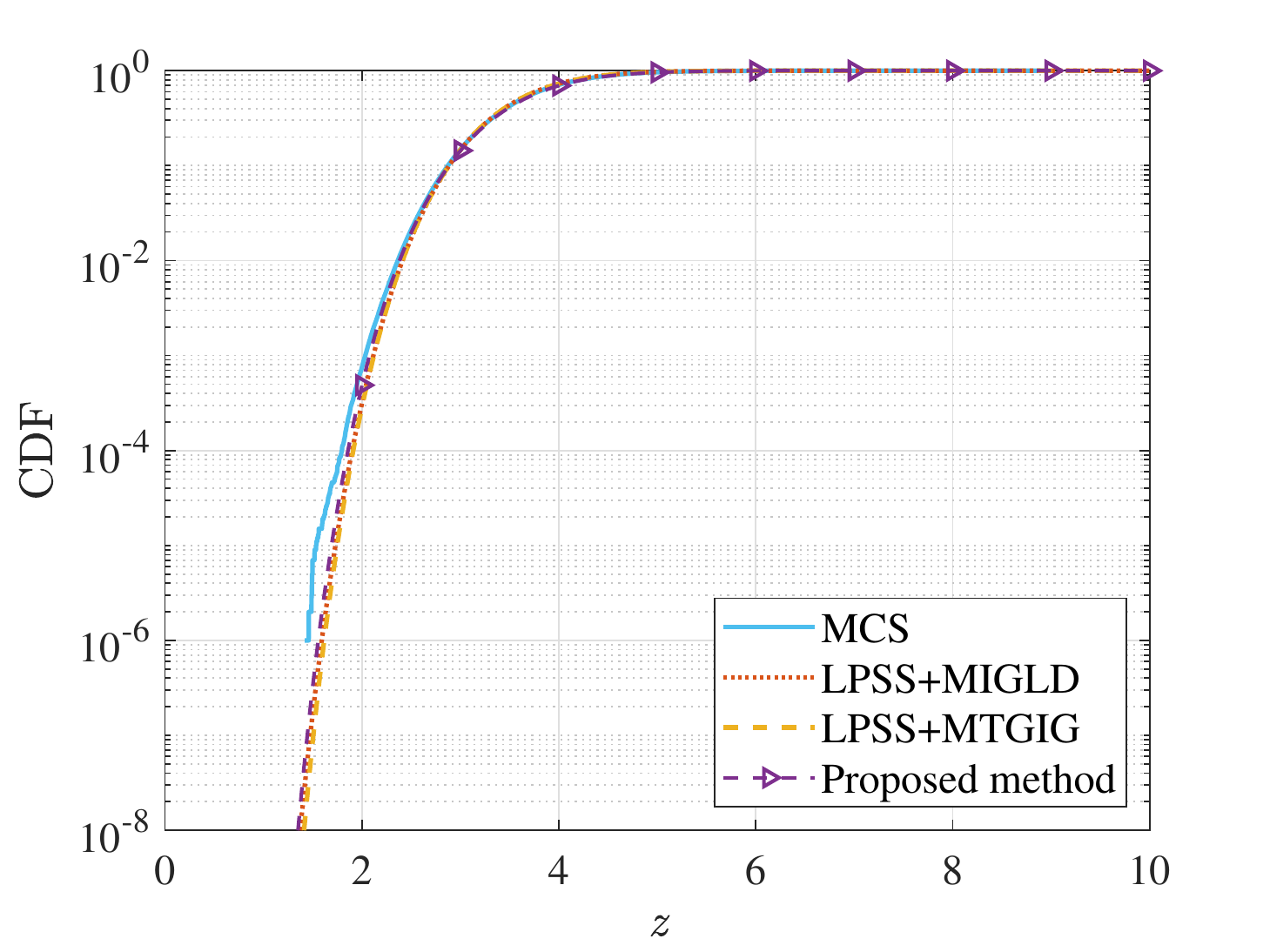}
		\end{minipage}
	}%
	
	\subfigure[POE]{
		\begin{minipage}{8.0cm}
			\centering
			\includegraphics[scale=0.50]{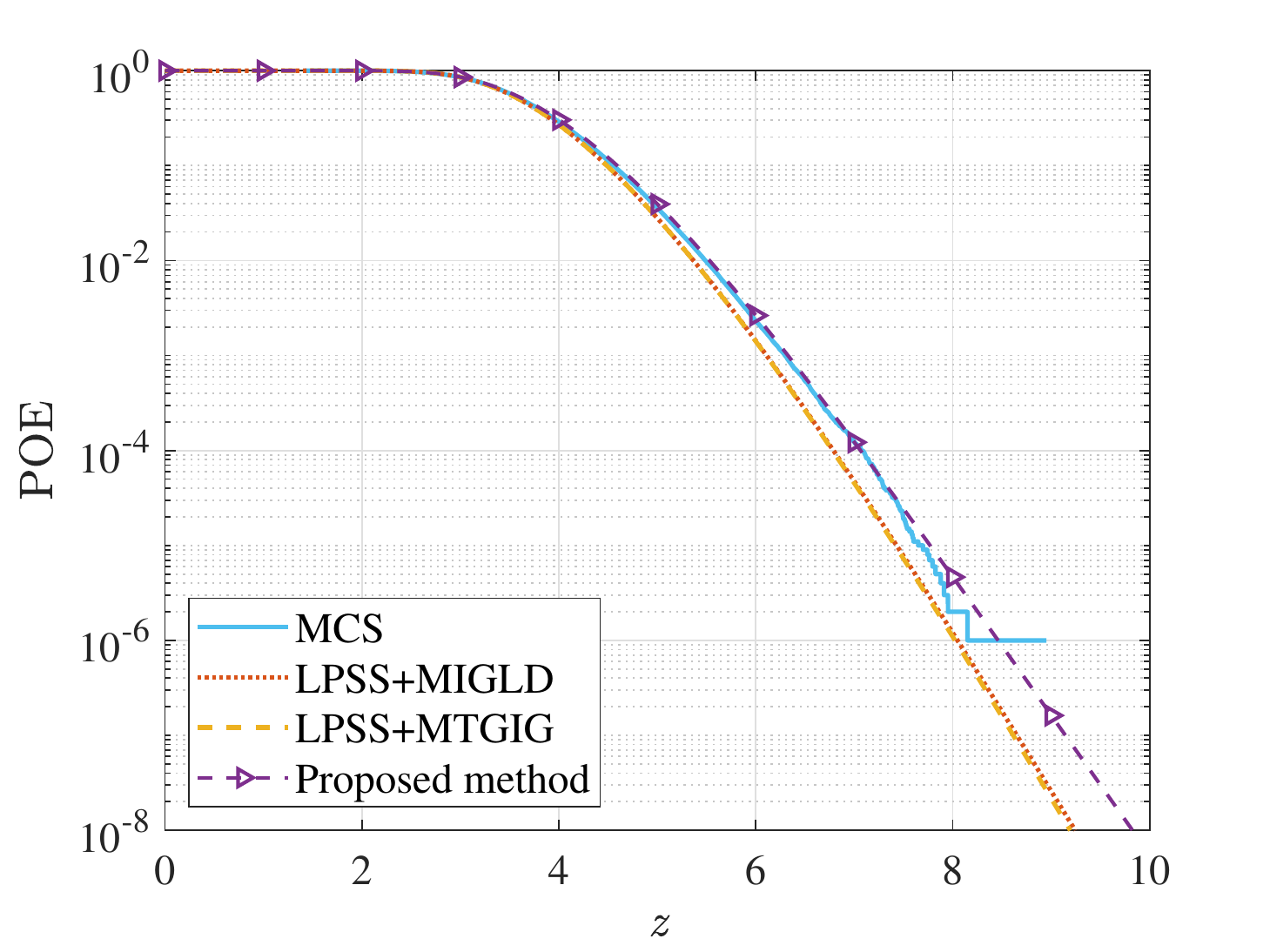}
		\end{minipage}
	}%
	
	\caption{PDF, CDF and POE of $\mathcal{Z}$ in Example 1}
	\label{fig:pdf_cdf_example_1}
\end{figure}

Once the required fractional moments are obtained, eight unknown free parameters involved in the proposed mixture distribution (i.e., \textcolor{black}{M-EIGD-LESND}) can be determined by the fractional moment matching technique. 
Specifically, the nonlinear system of equations in Eq. (\ref{eq:Nonlin_equations_para}) is solved according to Algorithm \ref{alg:Algorithm_2}, where initial values of free parameters are given to speed up the solving process.
Afterwards, the EVD could be approximated by the proposed mixture distribution model, where the PDF, CDF and probability of exceedance (POE) curves are all plotted in Fig. \ref{fig:pdf_cdf_example_1}. 
For comparison, the benchmark results by MCS and the results from LPSS+MIGLD and LPSS+MTGIG are also depicted in Fig. \ref{fig:pdf_cdf_example_1}. 
It can be found that both the PDF and POE curves obtained from the proposed method accord well with the MCS results.
Although there is almost no difference between the CDF curves obtained by proposed method and those by existing \textcolor{black}{mixture distribution} methods, 
larger deviations appear in the POE curves obtained by the LPSS+MIGLD and LPSS+MTGIG. 
Moreover, both of the LPSS+MIGLD and LPSS+MTGIG require 625 LPSS samples to estimate the fractional moments used for distribution parameter evaluation, where the number of samples is empirically determined in advance and is larger than that required by the proposed method. 
In this regard, the proposed method shows a considerable improvement in both efficiency and accuracy to recover the EVD in this example. 

After obtaining the reconstructed EVD, the first-passage probability can be evaluated by Eq. (\ref{eq:MEVD2}), where the safe threshold of this example is set to be  $b_{\mathrm{lim}} = 7$. 
Table \ref{tab:Pf_comparison_example1} lists the first-passage probabilities estimated by the proposed method, LPSS+MIGLD, LPSS+MTGIG, SS and MCS.
In this table, the estimated \textcolor{black}{first-passage probabilities} are denoted as $\hat{P_f}$. 
With reference to \textcolor{black}{$\hat{P_f}$ obtained by the MCS}, i.e., $1.2200 \times 10^{-4}$, the first-passage probability evaluated by the proposed method has acceptable accuracy, which reads $1.2245 \times 10^{-4}$. 
Unfortunately, the first-passage probabilities by SS, LPSS+MIGLD and LPSS+MTGIG largely deviate from \textcolor{black}{the reference $\hat{P_f}$ by the MCS}. 

\begin{table}[htb!]
	\caption{Comparison of \textcolor{black}{first-passage probabilities} by different methods (Example 1)}
	\label{tab:Pf_comparison_example1}
	\centering 	
	\small
	\begin{tabular}{llllll}
		\toprule 
		Method	  & MCS  & SS & LPSS+MIGLD  & LPSS+MTGIG   & Proposed    \\
		\midrule       
		$\hat{\mathcal{N}}$     & $10^{6}$ & 4600 & 625  & 625  & 520    \\
		$\hat{P_f}$     & $1.2200 \times 10^{-4}$ & $8.3100 \times 10^{-5}$ & $4.7154 \times 10^{-5}$  & $4.5286 \times 10^{-5}$  & $1.2245 \times 10^{-4}$ \\
		
		\bottomrule       
	\end{tabular}
\end{table}


\subsection{Example 2: a 15-storey shear frame structure under fully nonstationary stochastic ground motion}
\begin{figure}[!htbp]
	\centering
	\includegraphics[width=0.5\linewidth]{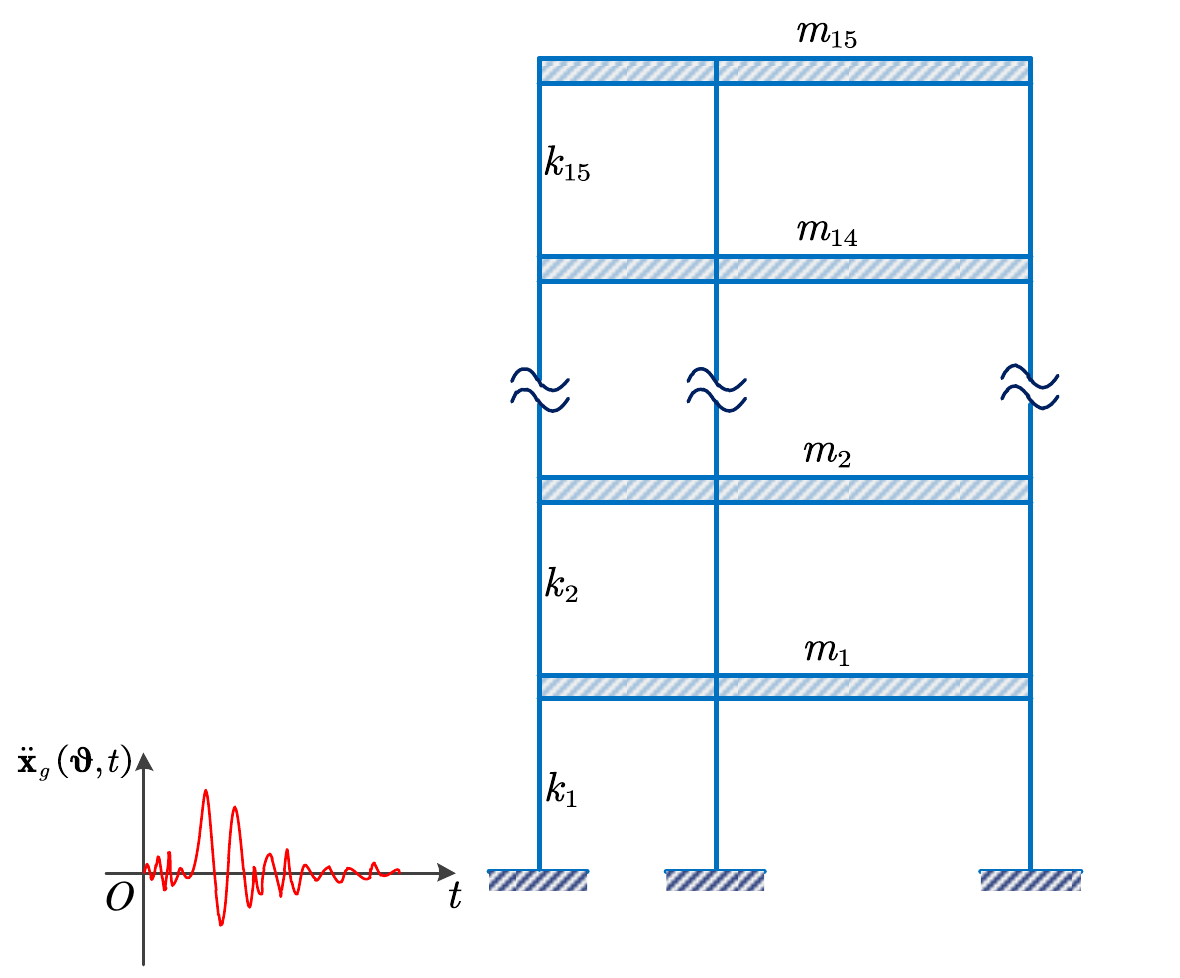}
	\caption{A 15-storey nonlinear shear frame structure}
	\label{fig:example_2}	
\end{figure}

A 15-storey nonlinear shear frame structure with uncertain structural properties under fully nonstationary stochastic ground motion is investigated in this example, shown in Fig. \ref{fig:example_2}. The equation of motion of this structure reads:
\begin{equation}\label{eq:equation_of_motion}
	\mathbf{M}\left( \mathbf{U }_{\mathrm{str}} \right) \mathbf{\ddot{Y}}+\mathbf{C}\left( {\mathbf{U }_{\mathrm{str}} } \right) \mathbf{\dot{Y}}+\mathbf{K}\left( {\mathbf{U }_{\mathrm{str}} } \right) \left[ \tilde{a}\mathbf{Y}+\left( 1-\tilde{a} \right) \mathbf{V} \right] =-\mathbf{M}\left( \mathbf{U}_{\mathrm{str}} \right) \mathbf{I\ddot{x}}_g\left( {\mathbf{U}_{\mathrm{exl}} },t \right), 
\end{equation}
where $\mathbf{\ddot{Y}}$, $\mathbf{\dot{Y}}$ and $\mathbf{{Y}}$ are the lateral acceleration, velocity and displacement matrices of the structure with respect to the ground; $\mathbf{M}$, $\mathbf{C}$ and $\mathbf{K}$ denote the mass, damping and stiffness matrices, respectively; Term $\mathbf{I}$ denotes the unit matrix. 
All of the lumped masses and the corresponding stiffnesses from bottom to top of the structure are assumed to be independent random variables, following the lognormal distributions with same coefficients of variation $0.1$ and different mean values $6 \times 10^4$ kg and $7 \times 10^7$ N/m, respectively. 
Hence, $n_{s_1}=30$ random variables are involved in the system properties, which are denoted as $\mathbf{U}_{\mathrm{str}}$. 
The floor slabs are assumed to be rigid. Rayleigh damping is implemented as $\mathbf{C}=\hat{\alpha}\mathbf{M}+\hat{\beta}\mathbf{K}
$, where $\hat{\alpha}$ and $\hat{\beta}$ are obtained by taking both the damping ratios of the first and second modes as 0.05. 
The Bouc-Wen resilience model \cite{wen1976method} is adopted to describe the nonlinear behavior of the structure, where the hysteretic displacement $\mathbf{V}$ satisfies:
\begin{equation}
\mathbf{\dot{V}}=\mathcal{A} \left( \varDelta \mathbf{\dot{Y}} \right) -\mathcal{B} \left| \varDelta \mathbf{\dot{Y}} \right|\left| \mathbf{V} \right|^{\rho -1}\mathbf{V}-\xi \left(\varDelta \mathbf{\dot{Y}}\right)\left| \mathbf{V} \right|^{\rho},
\end{equation}
in which $\varDelta \mathbf{\dot{Y}}$ is the relative velocity between two neighboring floors, $\tilde{a} = 0.1$, $\mathcal{A} = 1, \mathcal{B} = \xi = 50$ and $\rho = 1$ are the dimensionless parameters controlling the hysteretic performance of Bouc-Wen model. The fully nonstationary stochastic ground motion $\mathbf{\ddot{x}}_g\left( {\mathbf{U}_{\mathrm{exl}} },t \right)$ is modeled by the second family of spectral representation method (SRM) \cite{liu2016random}:
\begin{equation}\label{eq:time_history}
	\mathbf{\ddot{x}}_g\left( {\mathbf{U}_{\mathrm{exl}} },t \right) =\sqrt{2}\sum_{j=0}^{n_{s_2}-1}{\sqrt{2S_{\ddot{x}_g}\left( \omega _j,t \right) \varDelta \omega}}\cos \left( \omega _j t +{{U}_{\mathrm{exl},j}} \right),
\end{equation}
where ${\mathbf{U}_{\mathrm{exl}} } = \left[{{U}_{\mathrm{exl},1}},{{U}_{\mathrm{exl},2}},...,{{U}_{\mathrm{exl},n_{s_2}}}\right]$ denotes the random vector with $n_{s_2}=1600$ independent random variables uniformly distributed in $\left[0, 2\pi\right]^{n_{s_2}}$; $\omega_j = j\varDelta \omega, j=1,2,...,n_{s_2}$ is the discrete frequency and $\varDelta \omega = \omega_\mathrm{up}/{n_{s_2}}$ denotes the frequency interval with upper cut frequency $\omega_\mathrm{up}=240 \;\mathrm{rad/s}$;
${{S_{\ddot{x}_g}\left( \omega _j,t \right)}}$ is the double-sided evolutionary power spectrum density (EPSD) function:
\begin{equation}
	S_{\ddot{x}_g}\left( \omega ,t \right) =\left| \mathscr{A}\left( \omega ,t \right) \right|^2S\left( \omega \right), 
\end{equation}
in which $\mathscr{A}\left( \omega ,t \right)$ is the time-frequency modulation function and $S\left( \omega \right)$ is the power spectrum density represented by Clough-Penzien spectrum \cite{clough1975structures}, which are given as
\begin{equation}
	\mathscr{A}\left( \omega ,t \right) =e^{-\chi_0\frac{\omega t}{\omega _gT}}\cdot \left[ \frac{t}{\mathscr{C}_0}\cdot e^{\left( 1-\frac{t}{\mathscr{C}_0} \right)} \right]^{\kappa}, 
\end{equation}
\begin{equation}
	S\left( \omega \right) =\frac{\left[ \omega _{g}^{4}+4\zeta _{g}^{2}\omega _{g}^{2}\omega ^2 \right] \omega ^4}{\left[ \left( \omega _{g}^{2}-\omega ^2 \right) ^2+4\zeta _{g}^{2}\omega _{g}^{2}\omega ^2 \right] \left[ \left( \omega _{f}^{2}-\omega ^2 \right) ^2+4\zeta _{f}^{2}\omega _{f}^{2}\omega ^2 \right]}\frac{\bar{a}_{\max}^{2}}{\gamma _{0}^{2}\left[ \pi \omega _g\left( 2\zeta _g+\frac{1}{2\zeta _g} \right) \right]},
\end{equation}
where $\chi_0$ is the frequency modulation factor; $\mathscr{C}_0$ is the approximate arrive time of peak ground acceleration (PGA); $\kappa$ is the shape control coefficient; $\omega_g$ and $\zeta_g$ are the parameters describing the dominant frequency and damping ratio of site soil; $\omega_f$ and $\zeta_f$ are similar parameters for the second filter that hinders the low-frequency component; $\gamma_0$ is the peak factor; $T$ is the time duration; and $\bar{a}_{\max}$ denotes the PGA. 
Values of these involved parameters in EPSD take $\chi_0 = 0.15$, $\mathscr{C}_0 = 9 \;\mathrm{s}$, $\kappa = 2$, $\omega_f=0.1 \omega_g=\frac{4}{7}\pi$, $\zeta_f=\zeta_g=0.64$, $\gamma_0=2.85$, $T=20\;\mathrm{s}$, $\bar{a}_{\max}=400 \;\mathrm{cm/s^2}$. 
Note that a total number of $n_{s_1}+n_{s_2}=1630$ random variables are involved in this example. 

The maximum absolute extreme value of inter-storey drift on each storey over the time duration is considered as the response of interest in this example, which is denoted as $\mathcal{Z} _i, i=1,2,...,15$. 
Function solver \textit{Ode45} in Matlab is employed to perform deterministic dynamic analysis.
Here, the second-order fractional moment of the maximum value of $\mathcal{Z}$ of all layers, i.e., $\hat{M}_{\mathcal{Z}_{\mathrm{max}}}^2, \mathcal{Z} _{\mathrm{max}}=\underset{1\leqslant i\leqslant 15}{\max}\left\{ \mathcal{Z} _i \right\}$, is considered in the convergence criterion (Eq. (\ref{eq:convergence_criterion})). 
Accordingly, a total of $\hat{\mathcal{N}}=520$ samples of $\mathcal{Z} _i, i=1,2,...,15$ are generated, and the required fractional moments are obtained according to Eq. (\ref{eq:raw_moments_RLSS}).
\textcolor{black}{Besides, the speed up factor between the total computing time by using one CPU processor $T\left(1\right)$ and that by using 8 CPU processors $T\left(8\right)$ is computed, which is $S_p = T\left(1\right)/T\left(8\right) = 661 \;\rm{s}/246\;\rm{s} = 2.7$. 
This shows the benefit of using the parallel computing technique in the proposed parallel adaptive scheme.}

\begin{figure}[!htb]
	\centering
	\subfigure[PDF]{
		\begin{minipage}{8.0cm}
			\centering
			\includegraphics[scale=0.50]{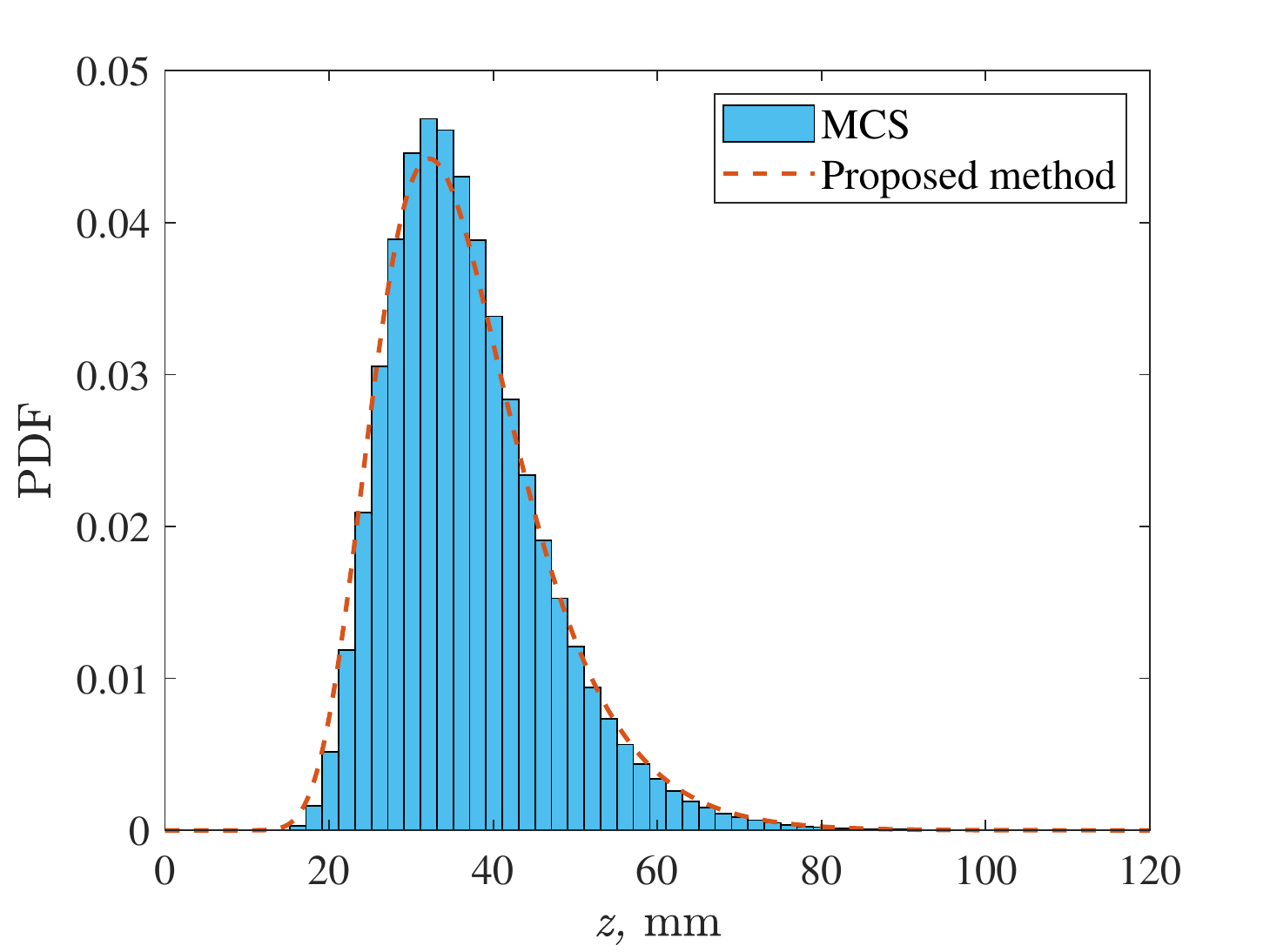}
		\end{minipage}
	}%
	\subfigure[POE]{
		\begin{minipage}{8.0cm}
			\centering
			\includegraphics[scale=0.50]{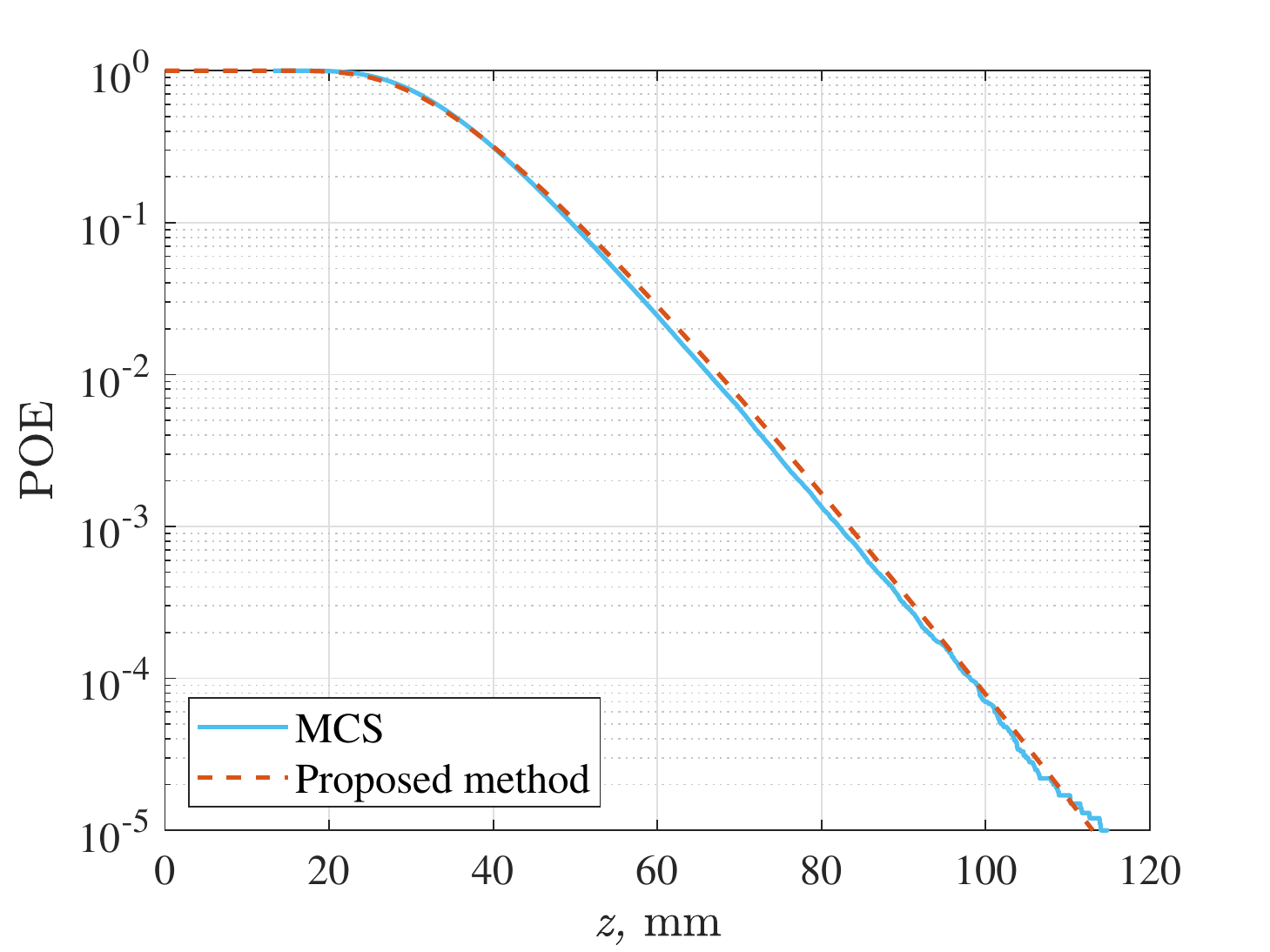}
		\end{minipage}
	}%
	\caption{PDF and POE of $\mathcal{Z}_1$ in Example 2}
	\label{fig:pdf_poe_1st_story_example_2}
\end{figure}

\begin{figure}[!htb]
	\centering
	\subfigure[PDF]{
		\begin{minipage}{8.0cm}
			\centering
			\includegraphics[scale=0.50]{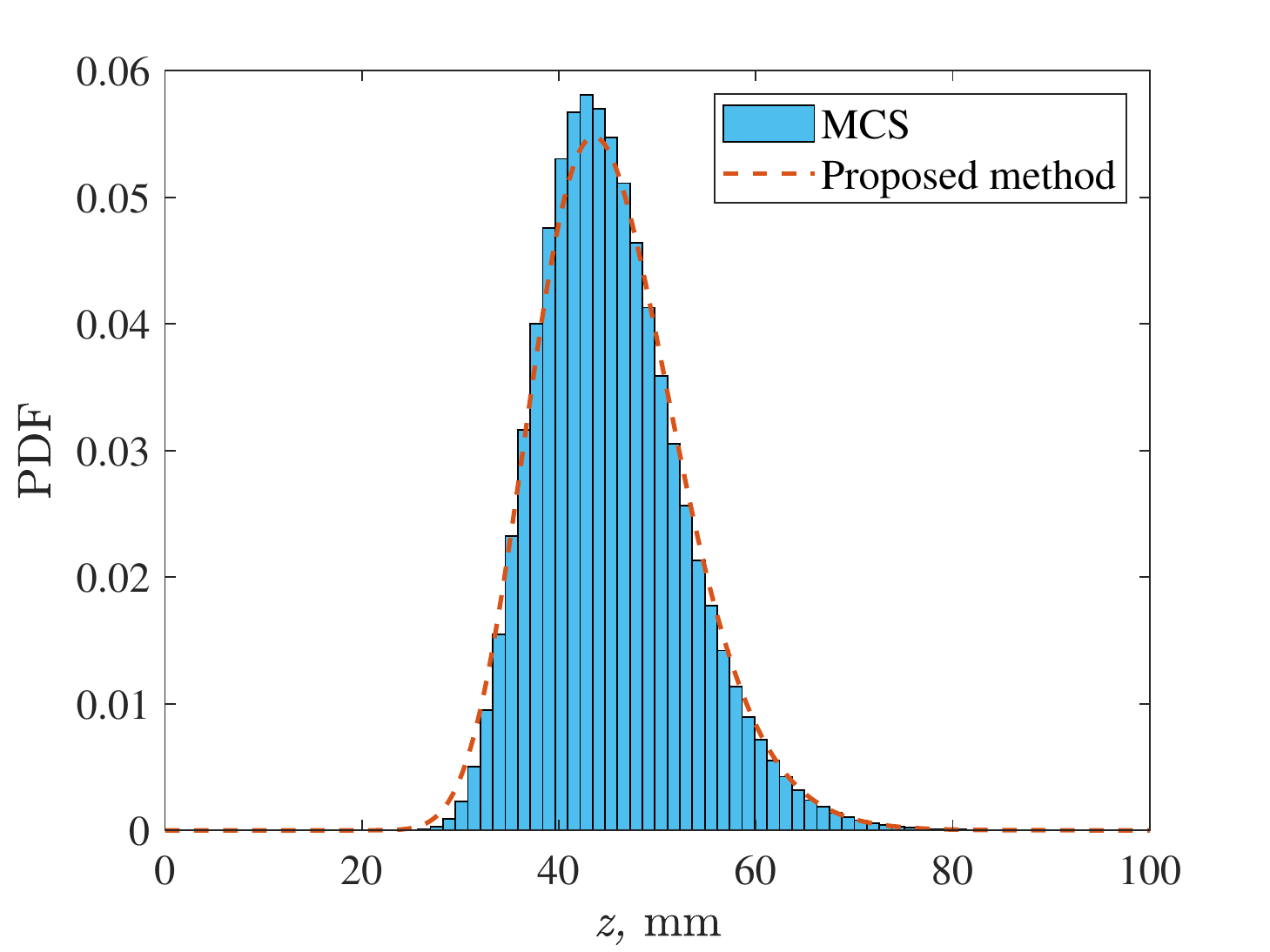}
		\end{minipage}
	}%
	\subfigure[POE]{
		\begin{minipage}{8.0cm}
			\centering
			\includegraphics[scale=0.50]{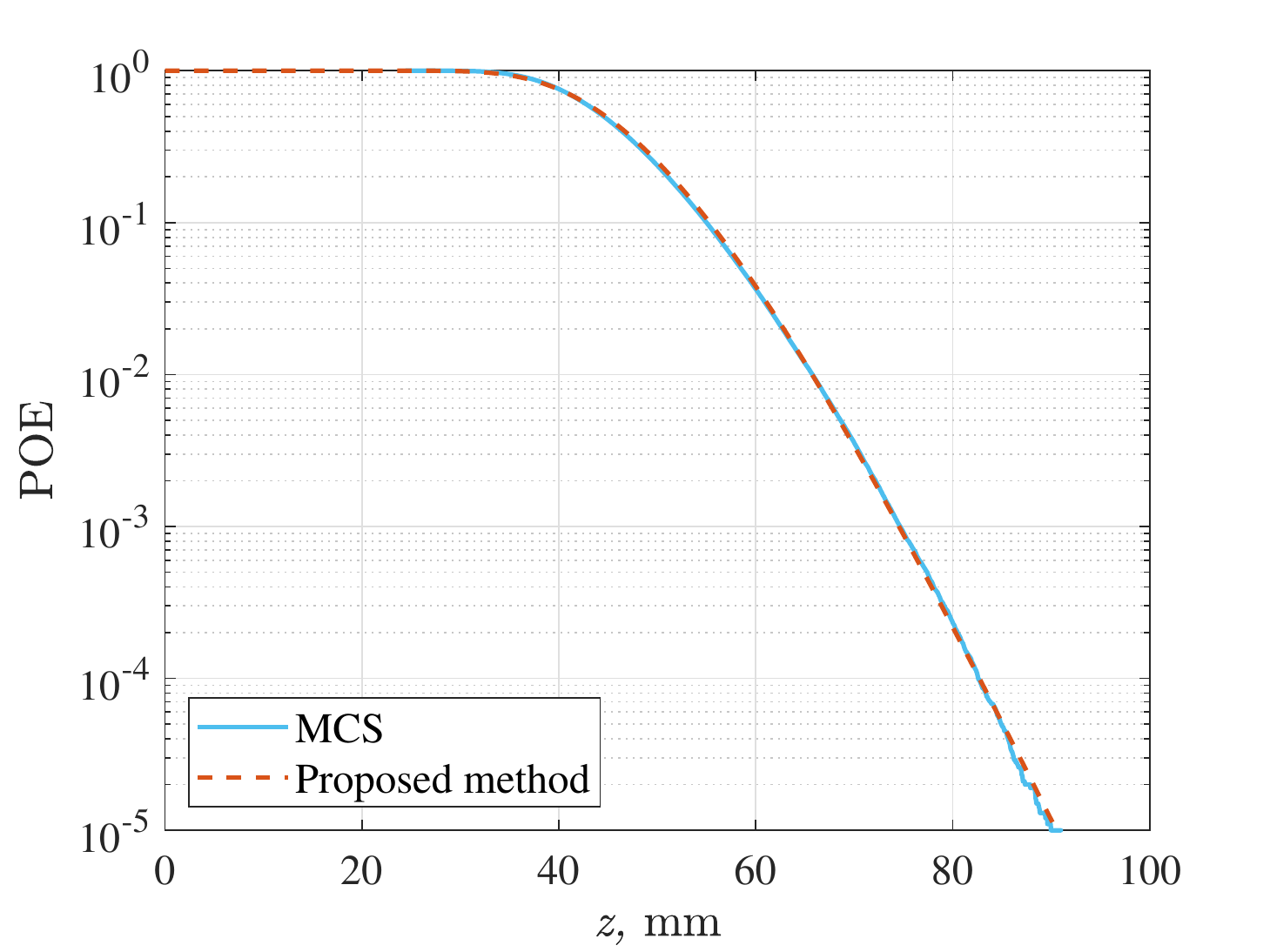}
		\end{minipage}
	}%
	\caption{PDF and POE of $\mathcal{Z}_7$ in Example 2}
	\label{fig:pdf_poe_8th_story_example_2}
\end{figure}

\begin{figure}[!htb]
	\centering
	\subfigure[PDF]{
		\begin{minipage}{8.0cm}
			\centering
			\includegraphics[scale=0.50]{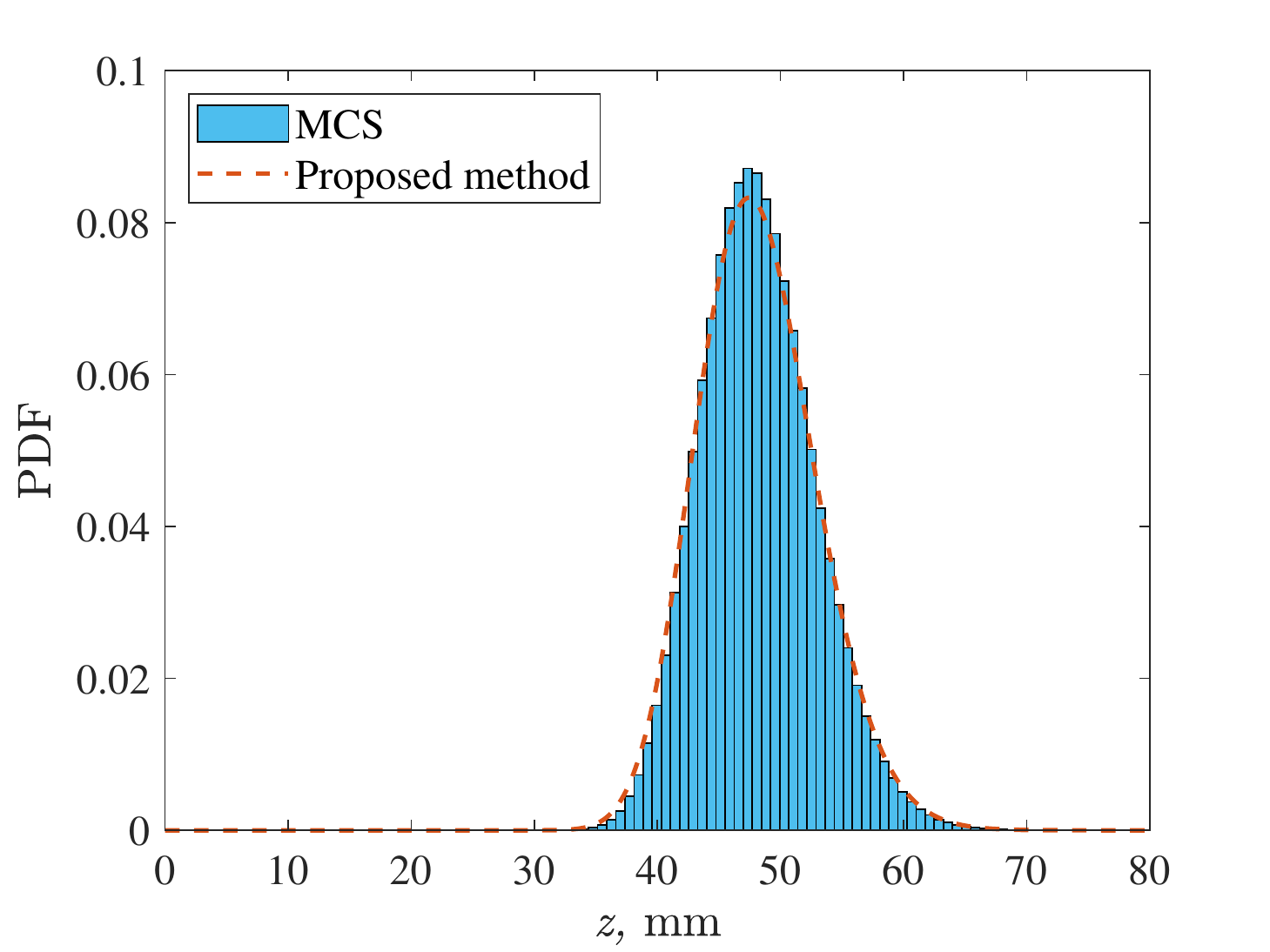}
		\end{minipage}
	}%
	\subfigure[POE]{
		\begin{minipage}{8.0cm}
			\centering
			\includegraphics[scale=0.50]{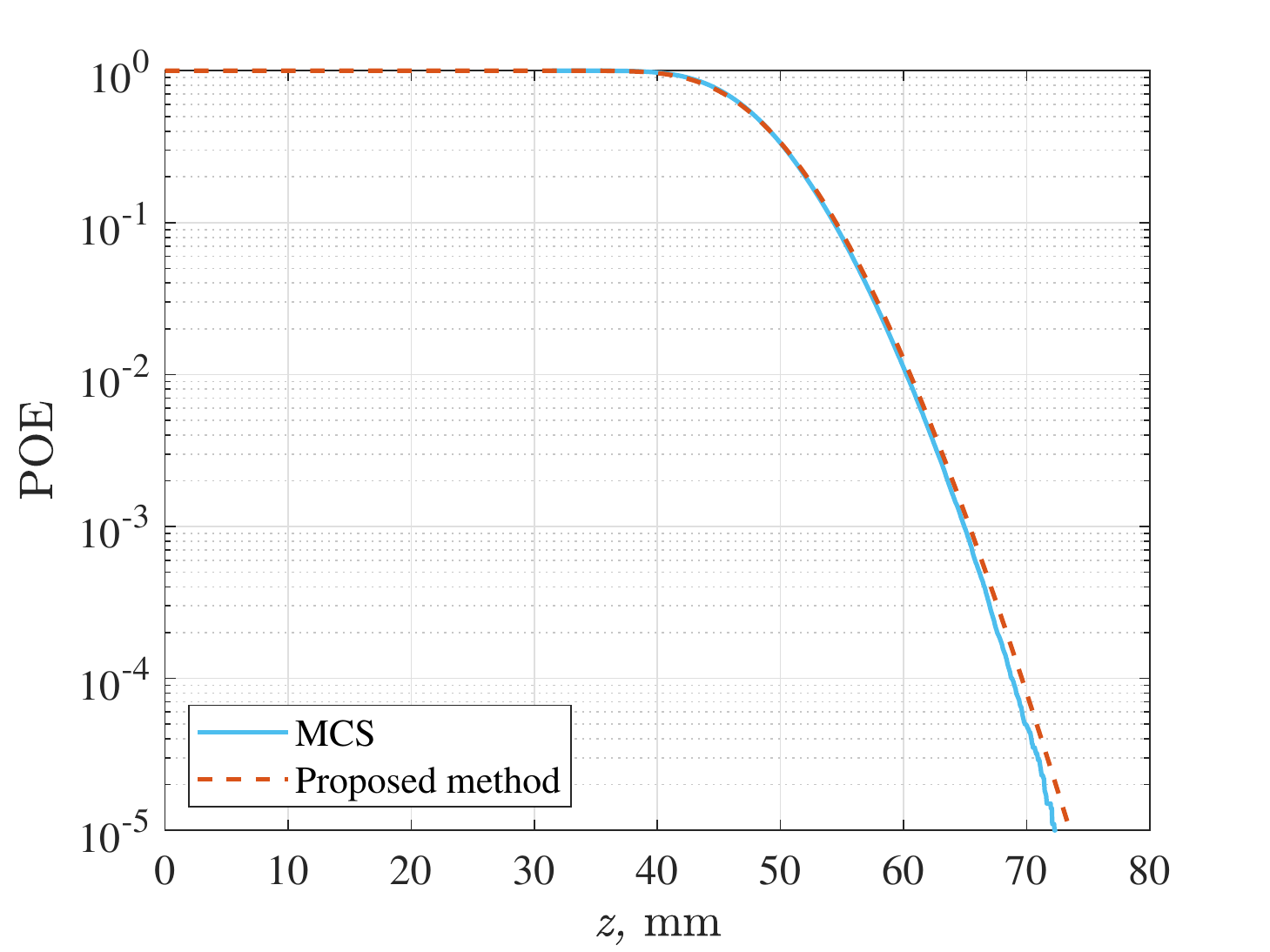}
		\end{minipage}
	}%
	\caption{PDF and POE of $\mathcal{Z}_{15}$ in Example 2}
	\label{fig:pdf_poe_15th_story_example_2}
\end{figure}

\begin{figure}[!htb]
	\centering
	\subfigure[PDF]{
		\begin{minipage}{8.0cm}
			\centering
			\includegraphics[scale=0.50]{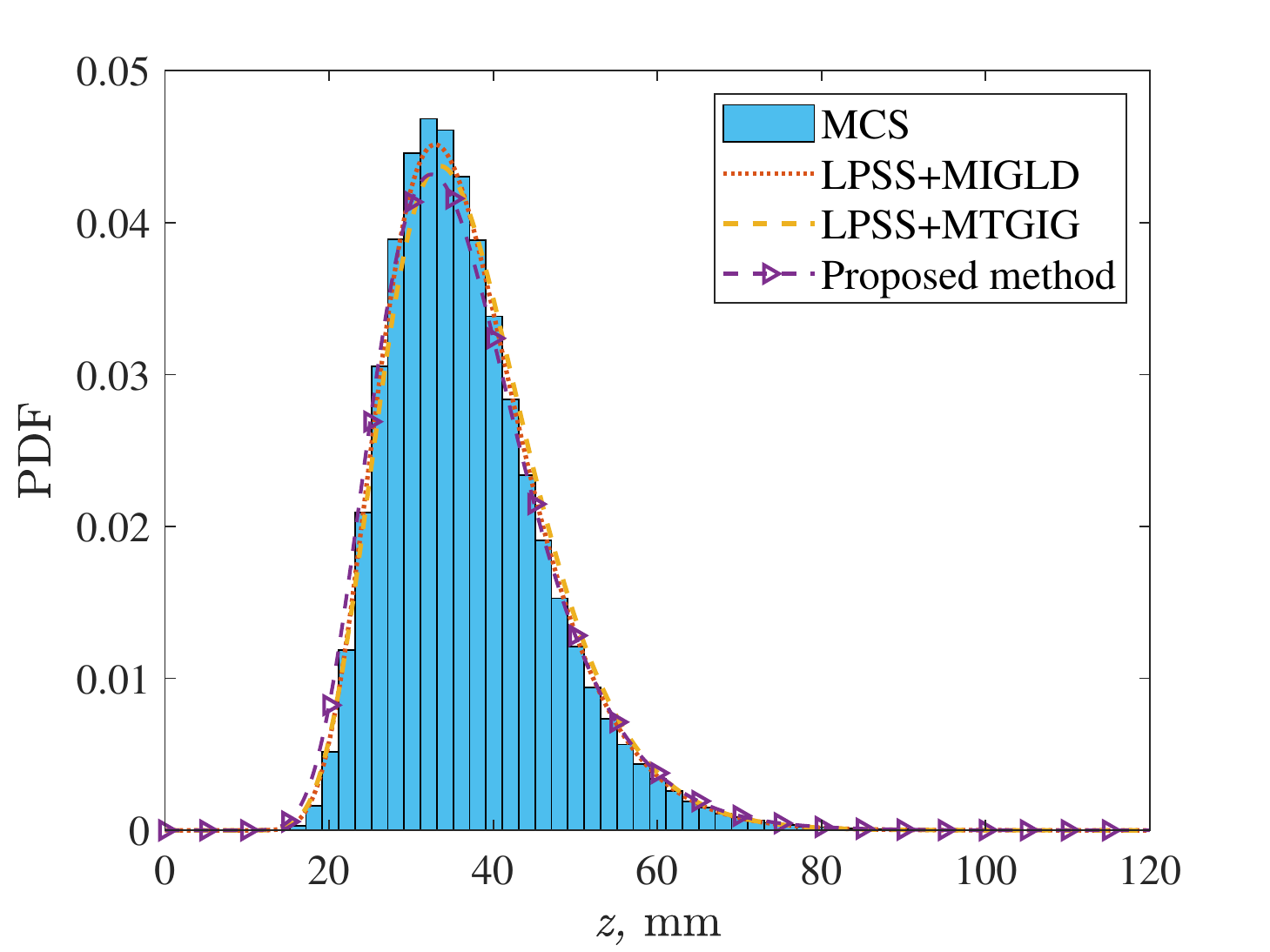}
		\end{minipage}
	}%
	\subfigure[POE]{
		\begin{minipage}{8.0cm}
			\centering
			\includegraphics[scale=0.50]{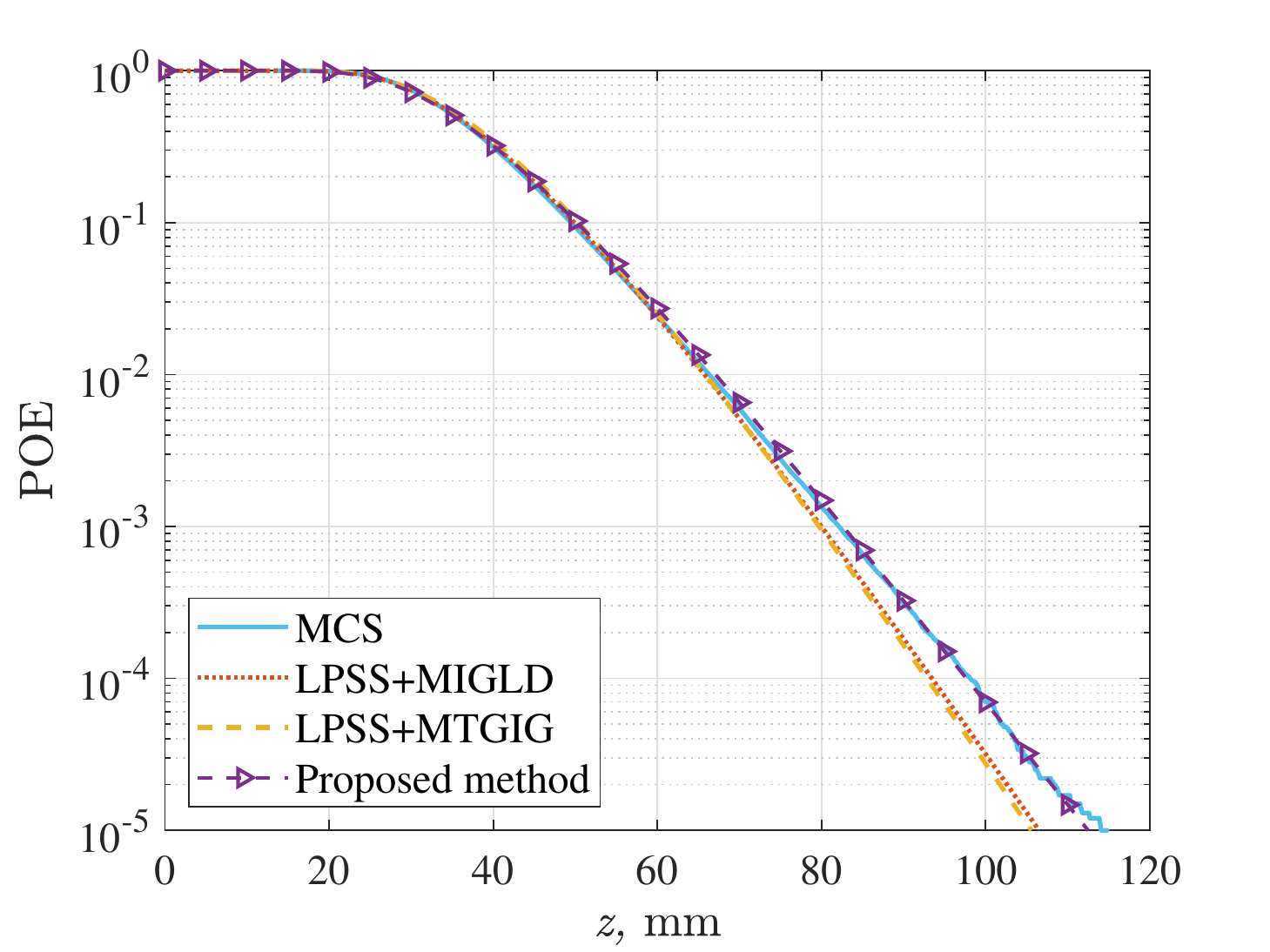}
		\end{minipage}
	}%
	\caption{A comparison between the PDF and POE of $\mathcal{Z}_{1}$ in Example 2}
	\label{fig:pdf_poe_15th_compared_story_example_2}
\end{figure}

Once the fractional moments are available, the EVDs of $\mathcal{Z} _i, i=1,2,...,15$ are then reconstructed by the proposed \textcolor{black}{M-EIGD-LESND}. 
Figs. \ref{fig:pdf_poe_1st_story_example_2}-\ref{fig:pdf_poe_15th_story_example_2} depict the PDFs and POEs of $\mathcal{Z}_1$ on the 1st storey, $\mathcal{Z}_7$ on the 7th storey and $\mathcal{Z}_{15}$ on the 15th storey, respectively. 
As seen, the proposed mixture distribution model well captures the main parts and tail information of the EVDs for selected storeys. 
Specifically, for all the selected storeys, the proposed method gives almost same accurate results of PDF and POE compared to the reference results from MCS. 
Besides, to further illustrate the advantages of the proposed method, a comparison of the PDF and POE curves of $\mathcal{Z}_{1}$ is depicted in Fig. \ref{fig:pdf_poe_15th_compared_story_example_2}, where results by LPSS+MIGLD and LPSS+MTGIG and those by the proposed method are given. 
As observed, with smaller sample size, the proposed method is able to capture the tail information more accurately than LPSS+MIGLD and LPSS+MTGIG, both of which require 625 samples. 

Further, we estimate the first-passage probabilities of the 1st, 7th and 15th storey of this example by Eq. (\ref{eq:MEVD2}), by setting three different thresholds as $b_{\mathrm{lim,1st}} = 95 \;\mathrm{mm}$, $b_{\mathrm{lim,7th}} = 80 \;\mathrm{mm}$ and $b_{\mathrm{lim,15th}} = 67 \;\mathrm{mm}$. 
Table \ref{tab:Pf_comparison_example2} gives the comparison results of proposed method, SS and MCS. 
As seen, with only 520 samples involved, all three first-passage probabilities by the proposed method have better accuracy than probabilities by SS. 

\begin{table}[!htb]
	\caption{Comparison of \textcolor{black}{first-passage probabilities} by the proposed method, SS and MCS (Example 2)}
	\label{tab:Pf_comparison_example2}
	\centering 	
	\begin{tabular}{ccccccc}
		\toprule
		\multirow{2}{*}{Method($\hat{\mathcal{N}}$)}  & \multicolumn{2}{c}{1st storey} &\multicolumn{2}{c}{7th storey} &\multicolumn{2}{c}{15th storey}\\
		\cline{2-7}
		& $b_{\mathrm{lim}}$(mm)   & $\hat{P_f}$ & $b_{\mathrm{lim}}$(mm)   & $\hat{P_f}$  & $b_{\mathrm{lim}}$(mm)   & $\hat{P_f}$              \\
		\midrule
		\textcolor{black}{M-EIGD-LESND}(520)  & 95   & $1.5075 \times 10^{-4}$  & 80   & $2.1708 \times 10^{-4}$   & 67   & $4.2208 \times 10^{-4}$         \\
		SS(3700)         & 95   & $1.9300 \times 10^{-4}$  & 80   & $4.3600 \times 10^{-4}$   & 67   & $4.5300 \times 10^{-4}$         \\
		MCS($10^6$)      & 95   & $1.6300 \times 10^{-4}$  & 80   & $2.3300 \times 10^{-4}$   & 67   & $3.0000 \times 10^{-4}$         \\
		\bottomrule                                                                            
	\end{tabular}
\end{table}

\subsection{ Example 3: a spatial steel frame structure with viscous dampers under fully nonstationary stochastic ground motion}
\begin{figure}[!htb]
	\centering
	\includegraphics[width=0.65\linewidth]{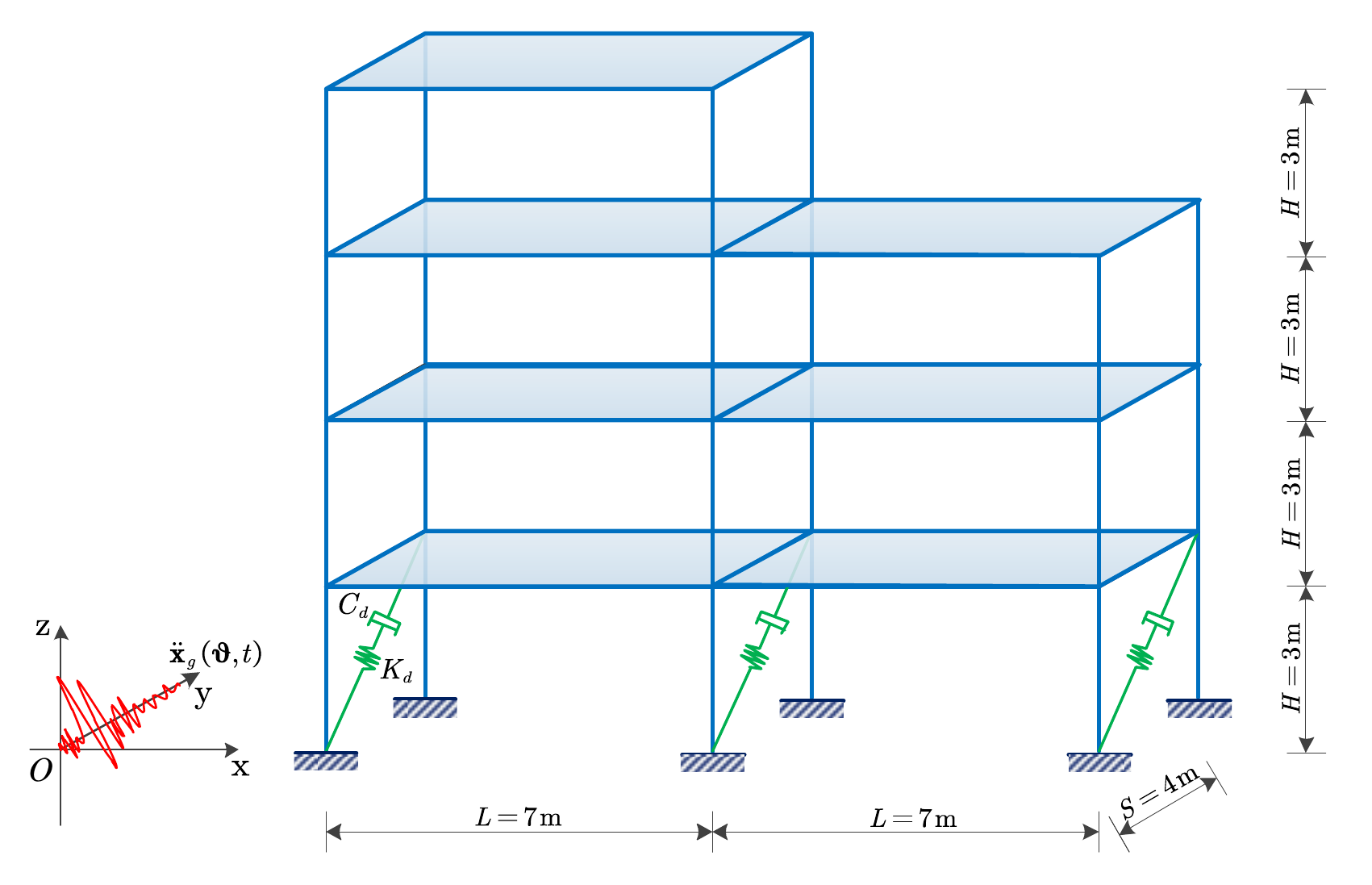}
	\caption{A two-bay four-storey nonlinear spatial steel frame structure with viscous dampers}
	\label{fig:example_3}	
\end{figure}

\begin{figure}[!htb]
	\centering
	\includegraphics[width=0.4\linewidth]{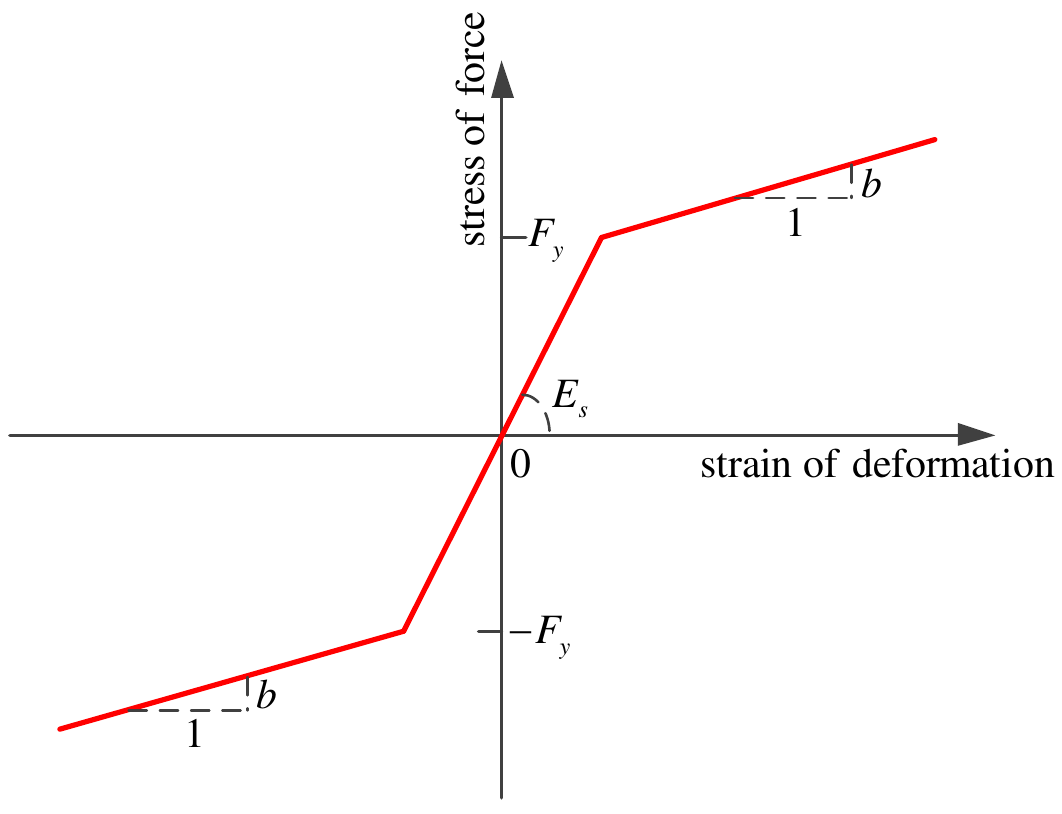}
	\caption{Bilinear constitutive model}
	\label{fig:steel01_example_3}	
\end{figure}

To illustrate the practical applicability of the proposed method, a two-bay four-storey nonlinear spatial steel frame structure with three viscous dampers under fully nonstationary ground motion is considered in this example, as shown in Fig. \ref{fig:example_3}. 
The whole structure is modeled and analyzed by the OpenSees software, where the bilinear constitutive model shown in Fig. \ref{fig:steel01_example_3} is used to model the nonlinear stress–strain relationship of steel materials. 
The slab of each floor is supposed to be rigid. 
The IPE270 beam and IPB300 column are adopted, where the column mass takes its self weight, while the beam mass is defined by ``self weight of beam + dead loads $D_L$ + 0.2 $\times$ live loads $L_L$". 
The viscous dampers are all represented by the Maxwell model which includes a linear spring and nonlinear dashpot in series. 
Three coefficients are involved in these viscous dampers, i.e., axial elastic stiffness of linear spring $K_d$, damping coefficient $C_d$, and velocity exponent $\alpha_d$. 
The Rayleigh damping is also employed here, where the damping ratios for both the first and second modes are taken as 0.03. 
The fully nonstationary stochastic ground motion takes the same form and parameters as employed in Example 2. 
It should be mentioned that the randomness of this structure \textcolor{black}{comes from} its external loads (i.e., dead loads, live loads and ground motion) and its structural properties. 
The statistical information of uncertain structural properties is collected in Table~\ref{tab:parameter_example3}. 
In total, 1608 random variables are involved in this example. 

\begin{table}[!htb]
	\caption{Statistical information of the uncertain structural properties in Example 3}
	\label{tab:parameter_example3}
	\centering	
	\begin{tabular}{lllll}
		\toprule
		Parameter	&	Description                  & Distribution  & Mean  & Standard variation          \\
		\midrule
		$D_L$	& Dead load                     & Lognormal & 10 $\mathrm{N/m^2}$ & 0.5 $\mathrm{N/m^2}$ \\
		$L_L$	& Live load                     & Lognormal & 10 $\mathrm{N/m^2}$ & 1 $\mathrm{N/m^2}$         \\
		$F_y$	& Yield strength of the steel   & Normal & $250\times 10^6$ Pa & $375\times 10^5$ Pa        \\
		$E_s$	& Young's modulus of the steel  & Normal & $2 \times 10^{11}$ Pa & $3 \times 10^{10}$ Pa \\
		$b$	    & Strain-hardening ratio        & Normal & $10^{-3}$ & $5 \times 10^{-5}$  \\
		$K_d$	& Axial stiffness of linear spring   & Normal & 25 Pa & 2.5 Pa    \\
		$C_d$	& Damping coefficient                 & Normal & 20.7452 & 2.07452   \\
		$\alpha_d$	& Velocity exponent             & Normal & 0.35 & 0.0175  \\
		\bottomrule
	\end{tabular}
\end{table}

\begin{figure}[!htb]
		\centering
		\subfigure[PDF]{
			\begin{minipage}{8.0cm}
				\centering
				\includegraphics[scale=0.50]{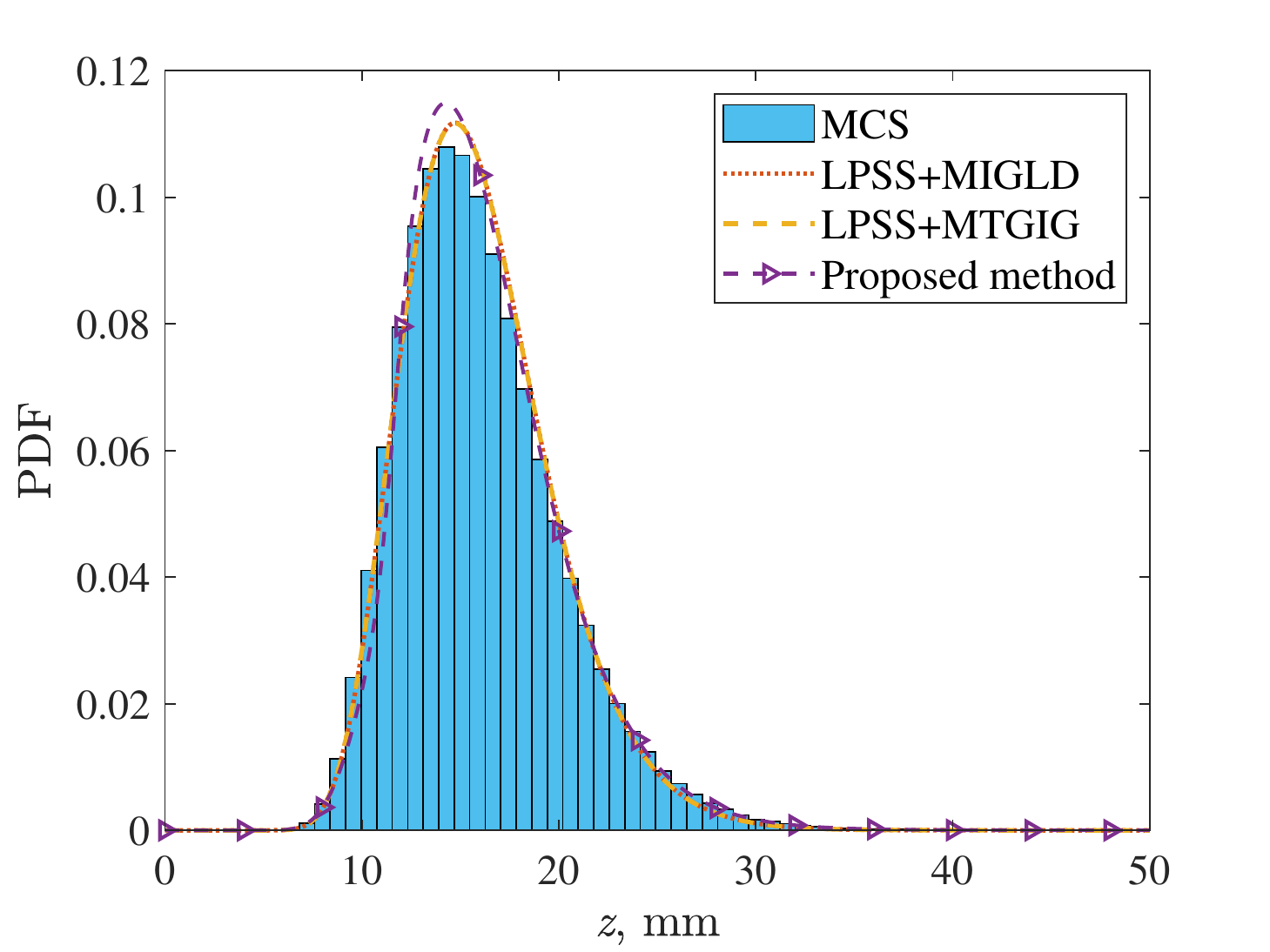}
			\end{minipage}
		}%
		\subfigure[POE]{
			\begin{minipage}{8.0cm}
				\centering
				\includegraphics[scale=0.50]{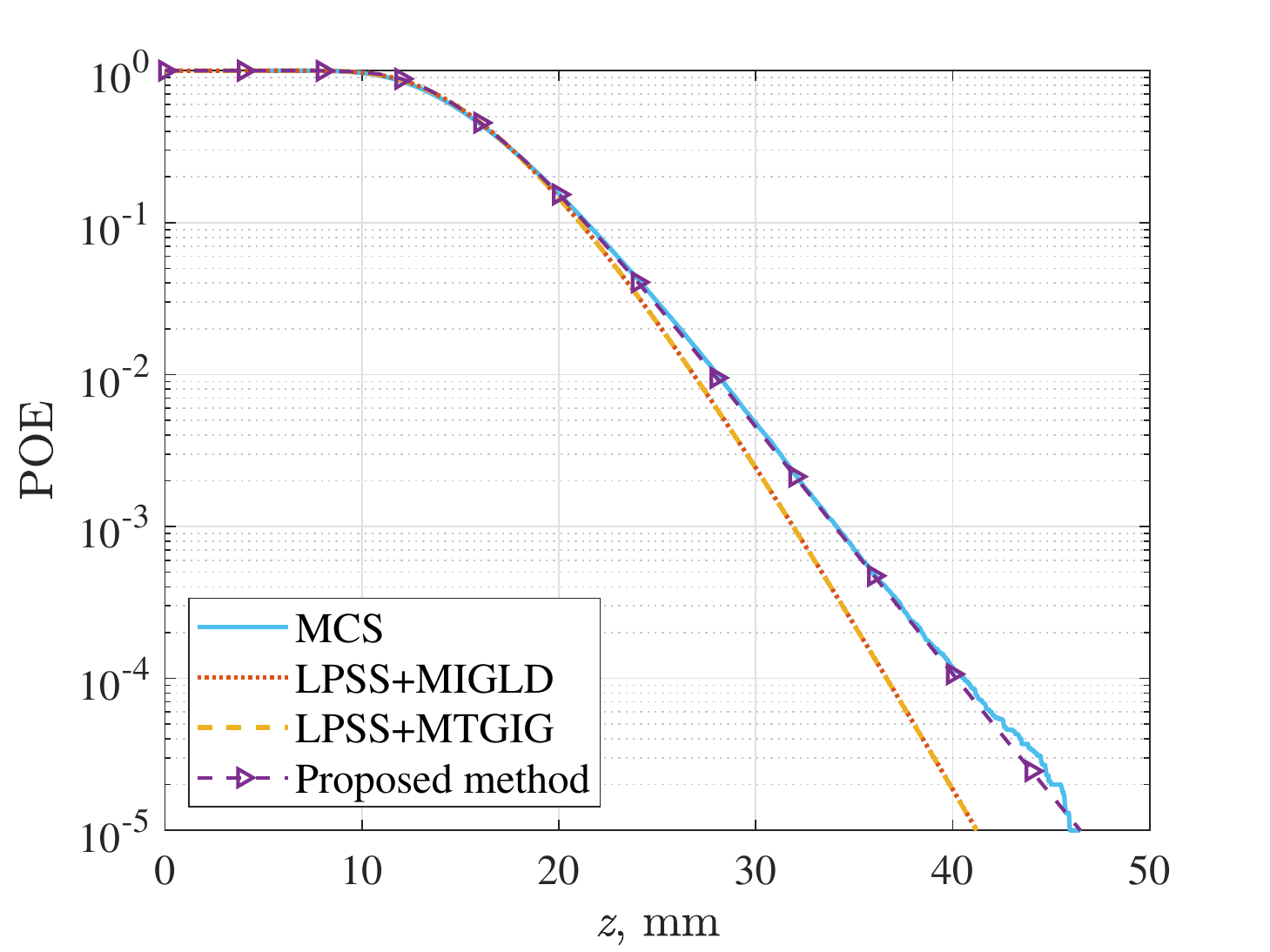}
			\end{minipage}
		}%
		\caption{PDF and POE of $\mathcal{Z}$ in Example 3}
		\label{fig:pdf_poe_largest_drift_example_3}
	\end{figure}

We consider the maximum absolute inter-storey drift of the whole structure as the quantity of interest, denoted by $\mathcal{Z}$. 
By adopting the proposed parallel adaptive scheme, $\hat{\mathcal{N}} = 1032$ samples of $\mathcal{Z}$ are generated, where a set of up to second order fractional moments can be estimated by Eq. (\ref{eq:raw_moments_RLSS}). 
From the knowledge of the estimated fractional moments, the EVD is represented by the proposed mixture distribution model, where the corresponding PDF and POE curves are depicted in Fig. \ref{fig:pdf_poe_largest_drift_example_3}. 
For comparison, the results by LPSS+MIGLD and LPSS+MTGIG are also provided, together with the benchmark results from MCS.
Good accordance between results by proposed method and MCS is readily observed. 
Admittedly, LPSS+MIGLD and LPSS+MTGIG are more computationally efficient since only 625 LPSS samples are employed.
However, the tail distributions captured by the LPSS+MIGLD and LPSS+MTGIG unfortunately deviate from the benchmark results to a large extent. 
Moreover, we calculate the first-passage probability of this example by setting the threshold of $\mathcal{Z}$ as 38 mm. 
\textcolor{black}{The first-passage probabilities by the MCS,} SS, LPSS+MIGLD, LPSS+MTGIG and proposed method are listed in Table \ref{tab:Pf_comparison_example3}. 
Remarkably, the proposed method yields a probability that is quite close to what MCS gives, i.e., $2.2439 \times 10^{-4}$ by the proposed method, and $2.3600 \times 10^{-4}$ by MCS. 
The probability by LPSS+MIGLD and LPSS+MTGIG notably deviate from the probability by the MCS, reading $5.0859 \times 10^{-5}$ and $5.0677 \times 10^{-5}$, respectively. 
In addition, the first-passage probability by SS is also less accurate, reading $2.0400 \times 10^{-4}$, but requires much more model evaluations.

	\begin{table}[htb]
		\caption{Comparison of \textcolor{black}{first-passage probabilities} by different methods (Example 3)}
		\label{tab:Pf_comparison_example3}
		\small
		\centering 	
		\begin{tabular}{llllll}
			\toprule 
			Method	  & MCS  & SS & LPSS+MIGLD  & LPSS+MTGIG   & Proposed   \\
			\midrule       
			$\hat{\mathcal{N}}$     & $  10^{6}$ & 3700 & 625  & 625  & 1032  \\
			$\hat{P_f}$     & $2.3600 \times 10^{-4}$  & $2.0400 \times 10^{-4}$  & $5.0859 \times 10^{-5}$  & $5.0677 \times 10^{-5}$  & $2.2439 \times 10^{-4}$  \\
			
			\bottomrule       
		\end{tabular}
	\end{table}
	
\section{Concluding remarks}\label{section:section5}
This paper proposes a novel fractional moments-based mixture distribution method to estimate the EVD and the first-passage probabilities of high-dimensional nonlinear stochastic dynamic systems. 
Unlike the existing methods, a parallel adaptive sampling scheme that allows for sample size extension is first proposed for estimating fractional moments.
By doing so, the sample size can be determined problem-dependently \textcolor{black}{in conjunction with} a proposed convergence criterion. 
Such scheme is realized by a sequential sampling method, i.e., refined latinized stratified sampling (RLSS), which also enables to achieve variance reduction in high dimensions. 
One versatile mixture distribution model, namely, \textcolor{black}{M-EIGD-LESND}, is proposed to represent the EVD with enhanced flexibility, whose free parameters are evaluated from obtained fractional moments. 
Three examples involving high-dimensional and strong-nonlinear stochastic dynamic systems are investigated to demonstrate the efficacy of the proposed method. 
The main conclusions are summarized as follows:
	
	(1) The studied examples indicate that the proposed method is able to tackle with \textcolor{black}{high-dimensional} and strongly nonlinear stochastic dynamic systems, where the uncertainties in both internal structural properties and external excitations are considered. 
	In addition, the proposed method is capable of accurately estimating small first-passage \textcolor{black}{probabilities} in the order of $10^{-4}$.
	
	(2) Several byproducts can be obtained by adopting the proposed method, i.e., fractional moments (including integer moments such as mean and standard deviation) and EVD. 
	Furthermore, for a general stochastic dynamic system, multiple EVDs \textcolor{black}{ and first-passage probabilities under different thresholds} can be estimated from only a single run of the proposed method. 
	
	(3) The proposed method is computational efficient since the proposed parallel adaptive scheme allows to determine an optimal sample size for a particular problem at hand. 
	In addition, only additional samples of extreme value need to be evaluated in each sample size extension, where parallel computing technique can be adopted to further improve the efficiency.
	
	(4) The proposed eight-parameter mixture distribution model is highly flexible and can adapt to different levels of distribution asymmetry. 
	This model generalizes several single-component distributions, such as the lognormal, skew-normal, log skew-normal, and inverse Gaussian distribution.
	In addition, \textcolor{black}{the mixture of lognormal and inverse Gaussian distributions is a special case of the proposed model.} 
	As a result, this model enables the proposed method to accurately recover a wide variety of EVDs. 

	\section*{CRediT authorship contribution statement}
	\textbf{Chen Ding:} Methodology, Software, Validation, Investigation, Writing - Original Draft, Writing - Revised draft; 
	\textbf{Chao Dang:} Conceptualization, Methodology, Investigation, Visualization, Writing - Original Draft, Writing - Revised draft, Funding acquisition; 
	\textbf{Marcos Valdebenito:} Validation, Writing- Reviewing and Editing, Funding acquisition; 
	\textbf{Matthias Faes:} Validation, Writing- Reviewing and Editing; 
	\textbf{Matteo Broggi:}  Validation, Supervision, Writing- Reviewing and Editing; 
	\textbf{Michael Beer:} Supervision, Project administration.
	
	\section*{Declaration of Competing Interest}
	The authors declare that they have no known competing financial interests or personal relationships that could have appeared to influence the work reported in this paper.
	
    \section*{Acknowledgement}
    Chen Ding is grateful for the support by the European Union’s Horizon 2020 research and innovation programme under Marie Sklodowska-Curie project GREYDIENT – Grant Agreement n°955393. Chao Dang is mainly supported by the China Scholarship Council (CSC). Marcos Valdebenito acknowledges the support of ANID (National Agency for Research and Development, Chile) under its program FONDECYT, grant number 1180271. 
	
\appendix
\section{Refined Latinized stratified sampling}\label{section:appendix}
To generate samples and weights by the Refined Latinized stratified sampling (RLSS) \cite{shields2016refined}, first we need to generate candidate samples and candidate strata by the combination of hierarchical Latin hypercube sampling (HLHS) \cite{shields2016refined} and Latinized stratified sampling (LSS) \cite{shields2016generalization}. 

Begin with a LSS design with $\mathcal{N}$ samples in $n_s$ dimensions.
First, generate a $n_s$-dimensional Latin hypercube sampling (LHS) design with $\mathcal{N}$ one-dimensional LHS strata $\Omega_{ij}$ and samples in each stratum $\varphi_{ij}$, $i=1,...,n_s; j=1,...,\mathcal{N}$. 
Denote $\mathcal{S}$ as the $\left[ 0,1 \right]^{n_s}$ space. 
Divide $\mathcal{S}$ equally into $\mathcal{N}$ mutually exclusive and collectively exhaustive strata ${\Omega } ^ {\left(k\right)}, k = 1,...,\mathcal{N}$, where ${\Omega } ^ {\left(k\right)} \bigcap{{\Omega } ^ {\left(q\right)}} = \emptyset, k \ne q$ and $\bigcup_{k=1}^{\mathcal{N}} {{\Omega }^{\left( k \right)}}=\mathcal{S}$. 
Note that each ${\Omega } ^ {\left(k\right)}$ is an equal-weighted hyper-rectangle and its boundary coincides with the boundary of $\Omega_{ij}$. 
Each ${\Omega } ^ {\left(k\right)}$ can be described by its starting coordinate near the origin $\boldsymbol{\varLambda}^{\left(k\right)} = \left\{\varLambda_{1}^{\left(k\right)},...,\varLambda_{n_s}^{\left(k\right)}\right\}$ and its side length $\lambda^{\left(k\right)} = \left\{\lambda_{1}^{\left(k\right)}, ... ,\lambda_{n_s}^{\left(k\right)}\right\}$. 
The weight of each ${\Omega } ^ {\left(k\right)}$ can be calculated as \cite{shields2016refined}:
\begin{equation}\label{eq:stratum_weight}
	\varpi ^{\left( k \right)}=\prod_{i=1}^{n_s}{\lambda _{i}^{\left( k \right)}},
\end{equation}
where $\sum_{k=1}^{{\mathcal{N}}}{\varpi ^{\left( k \right)}=1}$. 
For each ${\Omega } ^ {\left(k\right)}$, randomly pair each $\varphi_{ij}$ without replacement to produce the $k$-th LSS sample ${\boldsymbol{\varphi}} ^ {\left(k\right)} = \left[ {\varphi}_{1}^{\left(k\right)},...,{\varphi}_{n_s}^{\left(k\right)} \right], k = 1,...,\mathcal{N}$.

Afterwards, apply a $\delta$-level refinement of each $\Omega_{ij}$ based on the idea of HLHS, where $\delta \in \mathbb{Z}^{+}$ is the refinement factor. 
Specifically, along each dimension, divide $\Omega_{ij}$ $\delta$ times equally to obtain a total of $\tilde{\mathcal{N}} = \mathcal{N}\left(\delta+1\right)$ strata $\Omega_{ijh}, h=1,...,\tilde{\mathcal{N}}$.
Produce new candidate samples per each dimension by uniform sampling inside every empty newly produced stratum $\Omega_{ijh}$.
Subsequently, generate the candidate strata of RLSS, denoted as $\tilde{\Omega}^{\left(k^\star\right)}, k^\star=1,...,\tilde{\mathcal{N}}$, by dividing all the $\Omega^{\left({k}\right)}$ $\delta$ times along the LHS stratum boundaries in the dimension of largest side length $\lambda ^{\star} = \underset{i}{\max} \left\{ \lambda _{i}^{\left( k \right)} \right\}$. 
Then, identify the candidate stratum $\mathbf{\Xi }_{i}^{\left( k^\star \right)}=\left\{ \Omega _{ij}\in \left[ \varLambda _{i}^{\left( k^\star \right)},\varLambda _{i}^{\left( k^\star \right)}+\lambda _{i}^{\left( k^\star \right)} \right] \right\}$ which intersects with $\tilde{{\Omega }} ^ {\left(k^\star\right)}$ in each $i$-th dimension.
Count the number of $\mathbf{\Xi }_{i}^{\left( k^\star \right)}$ as $\varepsilon_{i}^{\left( k^\star \right)}, i=1,...,n_s$, and then determine the minimum number of $\mathbf{\Xi }_{i,\left( k^\star \right)}$ as ${\epsilon _i^\star}=\underset{k^\star}{\min}\left\{ \varepsilon _{i}^{\left( k^\star \right)} \right\}$. 
The candidate samples of RLSS, denoted as ${\tilde{\boldsymbol{\varphi}}}^{\left(k^\star \right)}, k^\star = 1,...,\tilde{\mathcal{N}}$, are generated by  drawing samples to the stratum $\Omega^{\left(k^\star\right)}$ satisfying $\varepsilon _{i}^{\left( k^\star \right)} = \epsilon _i^\star$: if $\epsilon _i^\star = 1$, $\Omega^{\left(k^\star\right)}$ contains only one single candidate LHS stratum, one must draw a sample from it; if $\epsilon _i^\star > 1$, one can draw samples from $\mathbf{\Xi }_{i}^{\left( k^\star \right)}$ at random without replacement. 
Repeat the sample adding process until all the dimensions of $\Omega^{\left(k^\star\right)}$ have one related sample. 

Once the candidate samples ${\tilde{\boldsymbol{\varphi}}}^{\left(k^\star \right)}$ and strata ${\tilde{{\Omega}}}^{\left(k^\star \right)}$ of RLSS are obtained, we can generate $\hbar$ RLSS samples at a time. 
First, randomly select $\hbar$ RLSS strata $\hat{\Omega}^{\left(l \right)}, l=1,...,\hbar$ from the candidate strata $\tilde{\Omega}^{\left(k^\star\right)}$. 
Then form RLSS samples ${\hat{\boldsymbol{\varphi}}}^{\left(l\right)}, l=1,...,\hbar$ by drawing corresponding samples from ${\tilde{\boldsymbol{\varphi}}}^{\left(k^\star \right)}$ to $\hat{\Omega}^{\left(l\right)}$. 
Update the stratum weight according to Eq. (\ref{eq:stratum_weight}) by specifying the side length of $\hat{\Omega}^{\left(l \right)}$.
Repeat several times to add $\hbar$ RLSS samples continuously until a user-defined convergence criterion is met or the number of remaining candidate samples ${\tilde{\boldsymbol{\varphi}}}^{\left(k^\star \right)}$ of RLSS is less than $\hbar$.  
Note that if the number of candidate samples is insufficient, a new extension of the sample candidate pool is required. 
If $\varsigma > 1$ \textcolor{black}{extensions of the candidate sample pool} can finally produce enough samples and weights of RLSS that meet the convergence criterion, then the total number of ${\tilde{\boldsymbol{\varphi}}}^{\left(k^\star \right)}$ and $\tilde{\Omega}^{\left(k^\star\right)}$ at this time will be $\tilde{\mathcal{N}} = \mathcal{N} \left( \delta +1 \right) ^{\varsigma}$. 
Briefly, the procedure of RLSS scheme is summarized in Algorithm \ref{alg:algorithm_1}, where $\hat{\mathcal{N}}$ denotes the obtained optimal sample size.

\begin{algorithm}[htb!]
	\caption{Refined Latinized stratified sampling approach \cite{shields2016refined}} 
	\label{alg:algorithm_1}
	\hspace*{0.02in} {\bf Input:} Dimension $n_s$ of the random parameter vector $\mathbf{U}$, LSS size $\mathcal{N}$, refinement factor $\delta$ and number of samples for each new sample size extension $\hbar$. 
	\\
	\hspace*{0.02in} {\bf Output:} 
	RLSS samples $\hat{\boldsymbol{\varphi}} = \left\{ \hat{\boldsymbol{\varphi}}^{\left(1\right)},...,\hat{\boldsymbol{\varphi}}^{\left(\hat{\mathcal{N}}\right)}  \right\}$ and corresponding weights ${\boldsymbol{\varpi}} = \left\{ {{\varpi}}^{\left(1\right)},...,{{\varpi}}^{\left(\hat{\mathcal{N}}\right)}  \right\}$.
	\begin{algorithmic}[1]
		\State Initialize with $\varsigma=1$. Define a LHS design with $\mathcal{N}$ ungrouped LHS sample components ${\varphi}_{ij}$ and corresponding one dimension LHS strata $\Omega_{ij}, i=1,...,n_s; j=1,...,\mathcal{N}$. 
		\State Establish a $n_s$-dimensional stratification ${\Omega }^{\left(k\right)}, k = 1,...,\mathcal{N}$ to form LSS strata such that each stratum is an equal-weighted hyper-rectangle and its boundary coincides with the boundary of ${\Omega}_{ij}$. Calculate the stratum weight of ${\Omega }^{\left(k\right)}$ according to Eq. (\ref{eq:stratum_weight}).
		\State Generate LSS samples ${\boldsymbol{\varphi}} ^ {\left(k\right)} = \left[ {\varphi}_{1}^{\left(k\right)} ,..., {\varphi}_{n_s}^{\left(k\right)} \right], k = 1,...,\mathcal{N}$ by randomly drawing ${\varphi}_{ij}$ to its related LSS stratum without replacement.
		\State Produce candidate samples per each dimension 
		by applying a $\delta$-level refinement of each ${\varphi}_{ij}$ inherent in ${{\Omega }}^{\left( k \right)}$ according to HLHS design.
		\State Generate candidate strata of RLSS $\tilde{\Omega}^{\left(k^\star\right)}, k^\star=1,...,{\mathcal{N}{\left( \delta+1 \right)^\varsigma}}$ by dividing all the strata ${\Omega } ^ {\left(k\right)}$ 
		equally $\delta$ times along every dimension with largest side length $\lambda_i^{\star}$. 
		\State Identify the strata $\mathbf{\Xi }_{i}^{\left( k^\star \right)}=\left\{ \Omega _{ij}\in \left[ \varLambda _{i}^{\left( k^\star \right)},\varLambda _{i}^{\left( k^\star \right)}+\lambda _{i}^{\left( k^\star \right)} \right] \right\}, k^\star=1,...,{\mathcal{N}{\left( \delta+1 \right)^\varsigma}}$ which intersect with $\tilde{\Omega}^{\left(k^\star\right)}$ in each $i$-th dimension.
		Count the number of $\mathbf{\Xi }_{i}^{\left( k^\star \right)}$ in the $i$-th dimension as $\varepsilon_{i}^{\left( k \right)}$, and then calculate ${\epsilon _i^\star}=\underset{k^\star}{\min}\left\{ \varepsilon _{i}^{\left( k^\star \right)} \right\}$.
		\State Generate candidate samples of RLSS ${\tilde{\boldsymbol{\varphi}}}^{\left(k^\star \right)}, k^\star = 1,...,{\mathcal{N}{\left( \delta+1 \right)^\varsigma}}$ inside the stratum $\tilde{\Omega}^{\left(k^\star\right)}$ satisfying $\varepsilon _{i}^{\left( k^\star \right)} = \epsilon _i^\star$: if $\epsilon _i^\star = 1$, draw samples from $\Omega_{ij}$; if $\epsilon _i^\star > 1$, draw samples from $\mathbf{\Xi }_{i}^{\left( k^\star \right)}$ at random; 
        repeat sample selection until all the dimensions are filled. 
		\State Select $\hbar$ RLSS strata $\hat{\Omega}^{\left(k \right)}, k=1,...,\hbar$ randomly from candidate $\tilde{\Omega}^{\left(k^\star\right)}$ and generate $\hbar$ RLSS samples ${\hat{\boldsymbol{\varphi}}}^{\left(k\right)}, k=1,...,\hbar$ by drawing corresponding samples from candidate ${\tilde{\boldsymbol{\varphi}}}^{\left(k^\star \right)}$ to $\hat{\Omega}^{\left(k\right)}$. Calculate the stratum weight according to Eq. (\ref{eq:stratum_weight}) by specifying the side length of $\hat{\Omega}^{\left(k \right)}$. 		
		\State  Repeat step 8 to add samples continuously until Eq. (\ref{eq:convergence_criterion}) is satisfied or an enlargement of the pool of candidate samples ${\tilde{\boldsymbol{\varphi}}}^{\left(k^\star \right)}$ is required. Then return to step 4 with $\varsigma=\varsigma+1$ and ${{\Omega }}^{\left( k \right)}=\tilde{\Omega}^{\left(k^\star\right)}$.
	\end{algorithmic}
\end{algorithm}

	\bibliographystyle{elsarticle-num} 
	\bibliography{reference}
	
	
	
	
	
\end{document}